\documentclass[aps,prd,onecolumn,showpacs,showkeys,amsmath,amssymb]{revtex4}
\usepackage{epsfig} 
\usepackage{amsmath}  
\usepackage{makecell}
\usepackage{graphicx}
\usepackage{array}
\usepackage{subfigure}
\usepackage{bm}
\usepackage{longtable}
\usepackage{booktabs}
\usepackage{appendix}
\usepackage{amsfonts}
\usepackage{amsmath}
\usepackage{dcolumn}
\usepackage{siunitx}
\usepackage{bm}
\usepackage{booktabs}
\usepackage[utf8]{inputenc}
\usepackage{threeparttable}
\usepackage{float}
\usepackage{longtable,lscape}
\usepackage{txfonts}
\usepackage{overpic}
\usepackage{amssymb}
\usepackage{indentfirst}
\usepackage{epsfig}
\usepackage{feynmf}   
\usepackage{epstopdf}   
\usepackage{slashed}  
\usepackage{multirow}
\usepackage[colorlinks, citecolor=blue,anchorcolor=red,menucolor=red, linkcolor=red,filecolor=red,runcolor=red,urlcolor=blue,frenchlinks=red]{hyperref}

\makeatletter
\newcommand{\figcaption}{\def\@captype{figure}\caption}
\newcommand{\tabcaption}{\def\@captype{table}\caption}

\newcommand{\Rmnum}[1]{\expandafter\@slowromancap\romannumeral #1@}
\def\hlinewd#1{%
  \noalign{\ifnum0=`}\fi\hrule \@height #1 \futurelet
   \reserved@a\@xhline}
\makeatother

\newcommand\qq{\langle\bar{q}q\rangle}
\newcommand\qss{\langle\bar{s}s\rangle}
\newcommand\GG{\langle GG\rangle}
\newcommand\qGq{\langle\bar{q}\sigma\cdot Gq\rangle}
\newcommand\sGs{\langle\bar{s}\sigma\cdot Gs\rangle}

\begin{document}
\title{The production and decay of $X_0(2900)$ state with different interpretation}
\author{Zi-Yan Yang$^{1,2}$}
\author{Qian Wang$^{1,2,4}$}
\email{qianwang@m.scnu.edu.cn}
\author{Wei Chen$^{3,4}$}
\email{chenwei29@mail.sysu.edu.cn}

\affiliation{$^1$Key Laboratory of Atomic and Subatomic Structure and Quantum Control (MOE), Guangdong Basic Research Center of Excellence for Structure and Fundamental Interactions of Matter, Institute of Quantum Matter, South China Normal University, Guangzhou 510006, China}
\affiliation{$^2$Guangdong-Hong Kong Joint Laboratory of Quantum Matter, Guangdong Provincial Key Laboratory of Nuclear Science, Southern Nuclear Science Computing Center, South China Normal University, Guangzhou 510006, China}
\affiliation{$^3$School of Physics, Sun Yat-sen University, Guangzhou 510275, China}
\affiliation{$^4$Southern Center for Nuclear-Science Theory (SCNT), Institute of Modern Physics, 
Chinese Academy of Sciences, Huizhou 516000, Guangdong Province, China}

\begin{abstract}
The observation of $X_0(2900)$ in $B^+\rightarrow D^+D^-K^+$ decay process indicates the existence of open flavor tetraquark states. We study the production and decay of $X_0(2900)$ state with final state interaction mechanism, where we calculate the strong vertices such as $g_{\bar{D}KX_0}$, $g_{\bar{D}^\ast K^\ast X_0}$, $g_{D_s^{\ast}\bar{D}K}$ and $g_{D_{s1}\bar{D} K^\ast}$ in the framework of QCD sum rules method. We find that for the interpretation of the $\bar{D}^\ast K^\ast$ molecule of $X_0(2900)$, the branching fraction of the production process and the decay width are consistent with the experimental results, indicating that the observed $X_0(2900)$ could be interpreted as the $\bar{D}^\ast K^\ast$ molecule. However, we cannot exclude the possibility of a compact tetraquark interpretation within the uncertainty. More experimental and theoretical efforts are needed to fully understand the nature of the $X_0(2900)$ state.
\end{abstract}
\pacs{12.39.Mk, 12.38.Lg, 14.40.Ev, 14.40.Rt}
\keywords{Tetraquark states, exotic states, hadron production, QCD sum rules}
\maketitle

\pagenumbering{arabic}

\section{Introduction}
\par The study of multiquarks goes back to 
half a century after the quark model proposed in 1964 ~\cite{Gell-Mann:1964ewy,1964-Zweig-p-} 
and becomes a hot topic since the observation of the $X(3872)$ in 2003. Due to the sufficient statistic in experiment, tens of charmonium-like and bottomonium-like states are observed by various experimental collaborations~\cite{ParticleDataGroup:2022pth}.
These hidden-charm(bottom) states could be beyond the conventional quark model and be viewed as exotic candidates.
They also provide a novel platform to
shed light on the hadronization mechanism.
Although numerous theoretical efforts have been put forward to understand their nature~\cite{Nielsen:2009uh,Chen:2016qju,Richard:2016eis,Esposito:2016noz,Ali:2017jda,Guo:2017jvc,Albuquerque:2018jkn,Liu:2019zoy,Brambilla:2019esw,Richard:2019cmi,Faustov:2021hjs,Chen:2022asf,Meng:2022ozq}, it is still unclear about the hadronization mechanism.  
\par Besides of the hidden-flavor structure, the exotic states can be composed of quarks with fully open flavors. The first fully open flavors tetraquark candidate was observed in 2016. The D0 collaboration observed a narrow structure $X(5568)$ in the $B_s^0\pi^\pm$ invariant mass spectrum with a 5.1$\sigma$ significance. $X(5568)$ could be interpreted as the $su\bar{d}\bar{b}$ or $sd\bar{u}\bar{b}$ tetraquark state due to the $B_s^0\pi^\pm$ decay mode~\cite{D0:2016mwd}. However, it was not confirmed by the LHCb~\cite{LHCb:2016dxl}, CMS~\cite{CMS:2017hfy}, CDF~\cite{CDF:2017dwr} and ATLAS~\cite{ATLAS:2018udc} collaborations. In 2020, the LHCb collaboration reported the observation of two new states $X_0(2900)$ and $X_{1}(2900)$ in the $D^-K^+$ invariant mass spectrum of the decay process $B^+\rightarrow D^+D^-K^+$~\cite{LHCb:2020bls,LHCb:2020pxc}. The resoncance parameters of these two new states are
\begin{equation}\label{Eq:ExpofX2900}
\begin{split}
 X_{0}(2900):&\;  m=2866\pm 7\pm 2 \mathrm{MeV},\\
             &\;  \Gamma=57\pm 12\pm 4 \mathrm{MeV},
\end{split}
\end{equation}
and
\begin{equation}
\begin{split}
 X_{1}(2900):&\;  m=2904\pm 5\pm 1 \mathrm{MeV},\\
             &\;  \Gamma=110\pm 11\pm 4 \mathrm{MeV},
\end{split}
\end{equation}
respectively. The fit fraction of $X_0(2900)$ in the process $B^+\rightarrow D^+D^-K^+$ is measured to be $(5.6\pm1.4\pm0.5)\%$ and the branching fraction is $(2.2\pm0.7)\times10^{-4}$. Thus the branching fraction of the two body cascade decay process $B^+\rightarrow D^+X_0(2900)\rightarrow D^+D^-K^+$ is measured to be $(1.2\pm0.5)\times10^{-5}$. Considering the isospin symmetry, one can approximately conclude that~\cite{Yu:2023avh}
\begin{equation}
\mathcal{B}r(B^+\rightarrow D^+X_0(2900))\sim (2.4\pm1.0)\times10^{-5}.
\end{equation}
Since they were observed in $D^-K^+$ channel, the minimal quark contents of these states should be $\bar{c}\bar{s}du$. Thus $X_{0,1}(2900)$ could be the strong candidates of fully open flavor tetraquark states and inspired theorists' extensive interest. 
\par Several studies indicate that $X_0(2900)$ state can be interpreted as $S$-wave $ud\bar{s}\bar{c}$ compact tetraquark state with different methods, such as chromomagnetic interaction model~\cite{Cheng:2020nho,Guo:2021mja,He:2020jna}, solving Schr\"odinger equation with variational method~\cite{Wang:2020prk} and QCD sum rule~\cite{Zhang:2020oze,Wang:2020xyc}. However, some theoretical studies give a negative results for the compact tetraquark interpretations of $X_0(2900)$, for example the extended relativized quark model~\cite{Lu:2020qmp}. In Ref.\cite{Agaev:2022eeh}, the authors revisit the result of Ref.\cite{Wang:2020xyc} and consider the decay properties in the framework of the light-cone sum rule. Their results disagree with the tetraquark interpretation for $X_0(2900)$. There are also some studies prefer the $\bar{D}^\ast K^\ast$ molecular interpretation for $X_0(2900)$ because its mass is very close to the threshold of $\bar{D}^\ast K^\ast$, such as the Lippmann-Schwinger equation with leading order contact potentials~\cite{Hu:2020mxp}, the Bethe-Salpeter framework~\cite{Kong:2021ohg,Ke:2022ocs}, one-boson exchange model~\cite{Liu:2020nil}, coupled-channel formalism~\cite{Wang:2021lwy} and QCD sum rule~\cite{Chen:2020aos,Agaev:2020nrc,Chen:2021erj,Mutuk:2020igv}. Even before the observation of the $X_0(2900)$, the $\bar{D}^*K^*$ molecular interpretation was investigated within hidden local gauge symmetry~\cite{Molina:2010tx} and the predicted mass was slightly lower than the observed one. The updated analysis within the same framework can be found in Ref.~\cite{Molina:2020hde}.
\par Several theoretical studies consider the decay and production properties of $X_0(2900)$~\cite{Huang:2020ptc,Xiao:2020ltm,Burns:2020xne,Burns:2020epm,Chen:2020eyu,Yu:2023avh,Hsiao:2021tyq}, which may be very sensitive to its inner structures of the multiquark system. In Ref.\cite{Huang:2020ptc}, the authors consider the two-body strong decays into $D^-K^+$ via triangle diagrams and three-body decays into $D^\ast \bar{K}\pi$ assuming that $X_0(2900)$ state is a bound state of $\bar{D}^{\ast}K^{\ast}$. They conclude that the $X_0(2900)$ may have a large $\bar{D}^{\ast}K^{\ast}$ component with a non-negligible compact tetraquark component. In Refs.\cite{Burns:2020xne,Burns:2020epm}, the authors make predictions for the production and decays of $X_0(2900)$ by modeling the production amplitude as a triangle diagram with non-perturbative final state interaction, and perturbative quark exchange, pion exchange and other effective field theory interactions. Their result shows that the triangle diagrams would also produce the $X_0(2900)$ signals, indicated that $X_0(2900)$ can be interpreted as a kinematic cusp effect arising from $\bar{D}^{\ast}K^{\ast}$ and $\bar{D}_1K^{(\ast)}$ interactions. Meanwhile, they do not exclude the resonance scenario. In Refs.\cite{Chen:2020eyu,Yu:2023avh}, the authors reproduce the branching fractions of $B^+\rightarrow D^+X_0$ via the rescattering mechanism.
\begin{figure}[htbp]
\centering
\includegraphics[width=8cm]{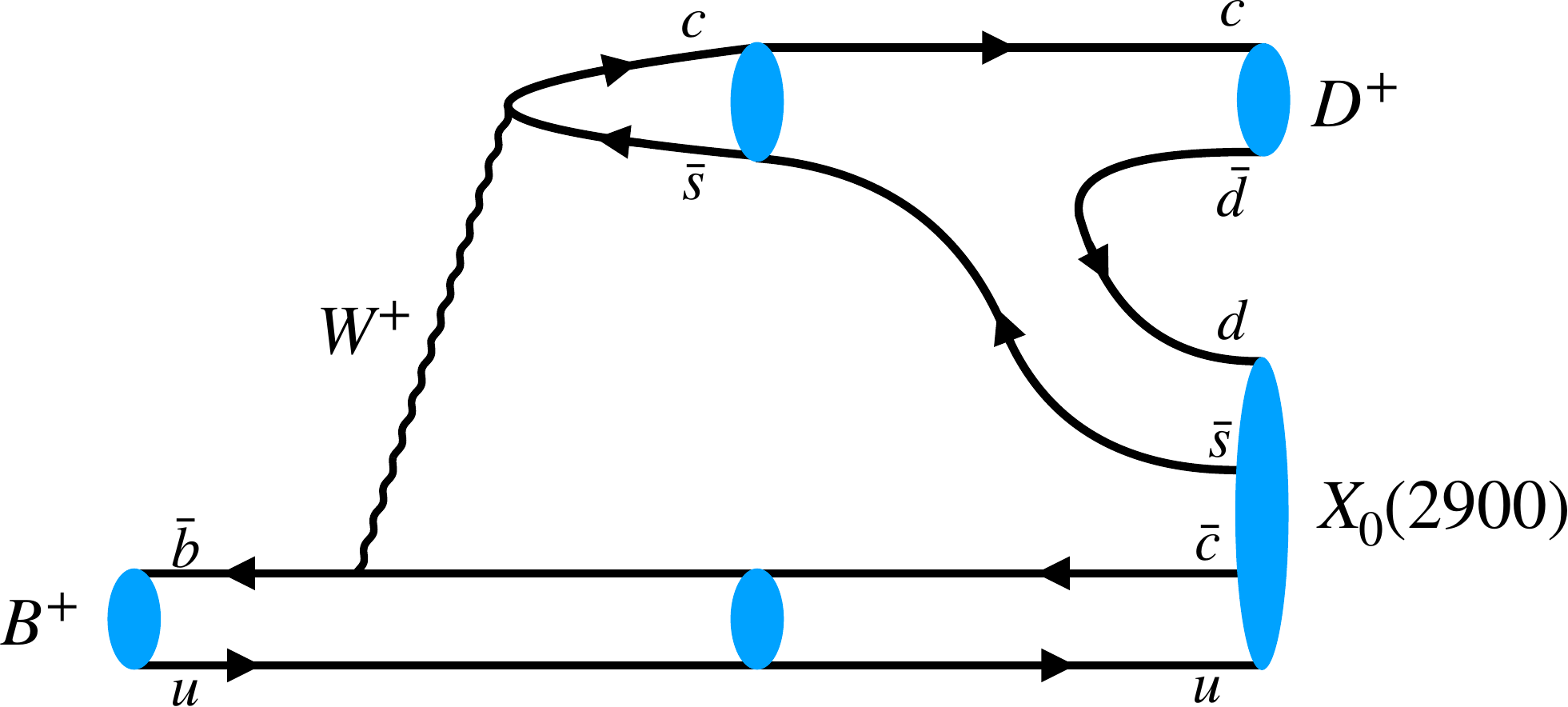}\\
\caption{Diagram contributing to $B^+\rightarrow D^+X_0(2900)$ at the quark level.}
\label{Fig:BDXquarklevel}
\end{figure}
\begin{figure}[htbp]
\centering
\includegraphics[width=\textwidth]{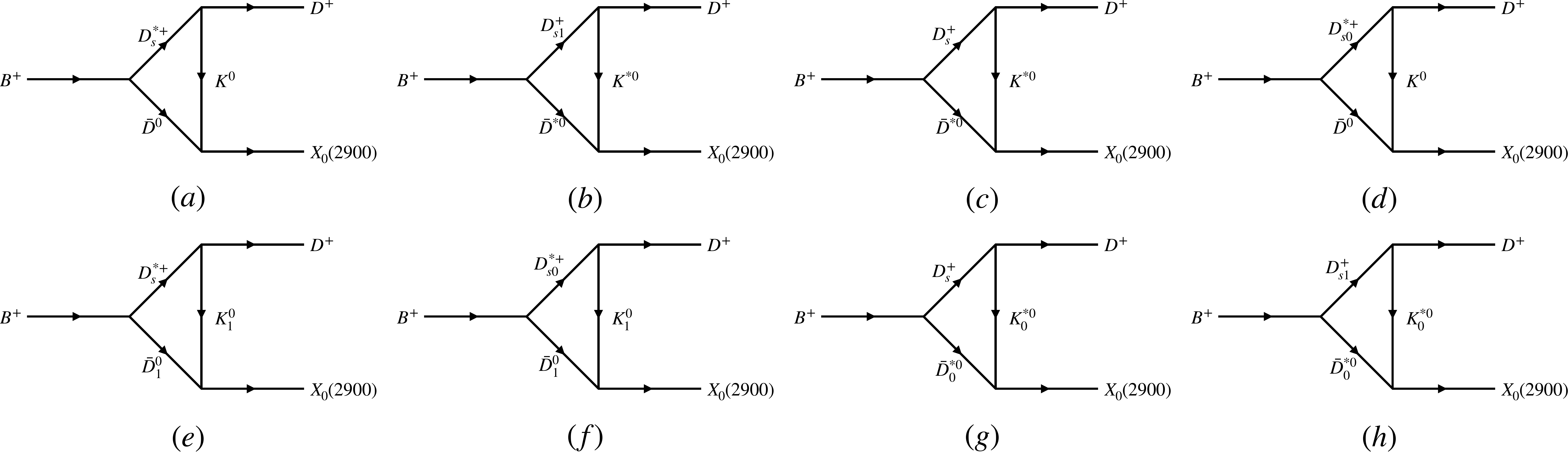}\\
\caption{Diagram contributing to $B^+\rightarrow D^+X_0(2900)$ at the hadron level.}
\label{Fig:BDXhadronlevel}
\end{figure}
\par At the quark level, the diagram for the process $B^+\rightarrow D^+ X_0(2900)$ is shown in Fig.~\ref{Fig:BDXquarklevel}. The weak decay arises from a Cabibbo-favored weak transition $\bar{b}\rightarrow\bar{c}(c\bar{s})$ along with the creation of a $d\bar{d}$ pair from the strong interaction. This diagram is the so-called external W emission diagram and it is not factorizable since the weak interaction produced $c\bar{s}$ enters into different final states. Thus, long-distance contributions would play a significant role in the $B^+\rightarrow D^+X_0(2900)$ process, where the weak interaction produced $c\bar{s}$ hadronize as $D_{s(1)}^{(\ast)+}$ and $u\bar{c}$ pair hadronize as $\bar{D}^{(\ast)0}$, followed by the strong rescattering in $D^+$ and $X_0(2900)$ with $K^0$ meson exchange, which is the process mainly concerned in~\cite{Chen:2020eyu}. Meanwhile in Ref.\cite{Yu:2023avh}, the authors consider the rescattering process $D^+_{s1}\bar{D}^{\ast0}\rightarrow D^+X_0(2900)$ with $K^{\ast0}$ meson exchange due to the fact that $X_0(2900)$ could be a good candidate for the $\bar{D}^\ast K^\ast$ molecule and mainly decays into $\bar{D}K$. The above two rescattering processes are shown in Fig.\ref{Fig:BDXhadronlevel}(a-b). Moreover, there should be other rescattering processes, such as $B\rightarrow D^\ast D_s\rightarrow D^+X_0(2900)$ with $K^\ast$ meson exchange. The other possible rescattering processes are shown in Fig.\ref{Fig:BDXhadronlevel}(c-h). The rescattering mechanism for the final-state-interaction (FSI) effects has been successfully applied to $D\rightarrow PV$ decays~\cite{Li:2002pj}, $B$ decays in $B\rightarrow\pi\pi,K\pi,D\pi$~\cite{Cheng:2004ru} and $KK$ channels~\cite{Lu:2005mx}. In this work, we will again test the rescattering mechanism in $B^+\rightarrow D_s^{\ast+}\bar{D}^0\rightarrow D^+X_0(2900)$ with $K^0$ exchange and $B^+\rightarrow D_{s1}\bar{D}^{\ast0}\rightarrow D^+X_0(2900)$ with $K^{\ast0}$ exchange, and other processes shown in Fig.~\ref{Fig:BDXhadronlevel}. 
In order to reveal the nature of the observed $X_0(2900)$, we consider $X_0(2900)$ as $\bar{D}^\ast K^\ast$ molecule structure, scalar diquark-scalar antidiquark structure and axial vector diquark-axial vector antidiquark structure with QCD sum rule approach in this work. 
\par This paper is organized as follows. In Sec.~\ref{Sec.FSI}, we derive the decay amplitudes for the  diagrams in Fig.~\ref{Fig:BDXhadronlevel} via final state interaction. In Sec.~\ref{Sec.QCDSR}, we use the QCD sum rules method to study the three-point correlation functions for strong coupling for $X_0(2900)$ such as $\bar{D}^0K^0X_0$, $\bar{D}^{\ast0}K^{\ast0}X_0$ and other strong coupling such as $D_{s}^{\ast+}D^+K^0$ and $D_{s1}^+D^+K^{\ast0}$ vertices. In Sec.~\ref{Sec.Production&Decay}, we will use the strong coupling constant obtained by QCD sum rules to calculate the amplitude and branching fraction for $B^+\rightarrow D^+X_0(2900)$ process. And finally we summarize our results in Sec.~\ref{Sec.Dis&Con}.
\section{Final state interaction}\label{Sec.FSI}
\par In the framework of rescattering mechanism, the decay $B^+\rightarrow D^+ X_0$ can most likely proceed as $B^+\rightarrow D_s^{\ast+}\bar{D}^0\rightarrow D^+X_0(2900)$ with $K^0$ exchange and $B^+\rightarrow D_{s1}\bar{D}^{\ast0}\rightarrow D^+X_0(2900)$ with $K^{\ast0}$ exchange~\cite{Chen:2020eyu,Yu:2023avh}, while other possible processes are also considered. Under the factorization approach~\cite{Wirbel:1985ji,Bauer:1986bm}, we can get the decay amplitude of $B^+\rightarrow D_s^{\ast+}\bar{D}^0$ and $B^+\rightarrow D_{s1}^+\bar{D}^{\ast0}$:
\begin{equation}\label{Eq:Factorization}
\begin{split}
\mathcal{A}(B^+\rightarrow D_s^{\ast+}\bar{D}^0)&=\frac{G_F}{\sqrt{2}}V_{cb}V_{cs}a_1\langle D_s^{\ast+}|J^{(c\bar{s})}_\mu|0\rangle\langle\bar{D}^0|J^W_\mu|B^+\rangle,\\
\mathcal{A}(B^+\rightarrow D_{s1}^+\bar{D}^{\ast0})&=\frac{G_F}{\sqrt{2}}V_{cb}V_{cs}a_1\langle D_{s1}^{+}|J^{(c\bar{s})}_\mu|0\rangle\langle\bar{D}^{\ast0}|J^W_\mu|B^+\rangle,\\    
\end{split}
\end{equation}
where $G_F$ is the Fermi constant, $V_{ik}$ is the CKM matrix elements, $a_1$ is the effective Wilson coefficients obtained by the factorization approach~\cite{Buchalla:1995vs}, and the effective operators $J^W_\mu, J^{(c\bar{s})_\mu}$ are defined as
\begin{equation}\label{Eq:EffOperator}
\begin{split}
J^W_\mu&=\bar{c}\gamma_\mu(1-\gamma_5)b,\\
J^{(c\bar{s})}_\mu&=\bar{s}\gamma_\mu(1-\gamma_5)c.
\end{split}
\end{equation}
The matrix elements in Eqs.\eqref{Eq:Factorization} are defined as:
\allowdisplaybreaks{
\begin{eqnarray}
\nonumber\langle 0|J^{(c\bar{s})}_\mu|D_s^{\ast+}(p_1,\epsilon)\rangle&=&\lambda_{D_s^{\ast+}}\epsilon_\mu, \\
\nonumber\langle 0|J^{(c\bar{s})}_\mu|D_{s1}^{+}(p_1,\epsilon)\rangle&=&-\lambda_{D_{s1}^{+}}\epsilon_\mu, \\
\langle\bar{D}^0(p_2)|J^W_\mu|B^+(p)\rangle&=&F_{+}(Q^2)P_\mu+F_{-}(Q^2)q_\mu, \\
\nonumber\langle\bar{D}^{\ast0}(p_2,\epsilon)|J^W_\mu|B^+(p)\rangle&=&\frac{\mathrm{i}\epsilon^\nu}{m_B+m_{D^\ast}}\Big\{\mathrm{i}\epsilon_{\mu\nu\alpha\beta}P^\alpha q^\beta A_V(Q^2)+(m_B+m_{D^\ast})^2g_{\mu\nu}A_1(Q^2)-P_\mu P_\nu A_2(Q^2)\\
\nonumber& &-2m_{D^\ast}(m_B+m_{D^\ast})\frac{P_\nu q_\mu}{q^2}\left(A_3(Q^2)-A_0(Q^2)\right)\Big\},
\end{eqnarray}}
where $P_\mu=p_\mu+p_{2\mu}, q_\mu=p_\mu-p_{2\mu}, Q^2=-q^2$ and $F_{\pm}(Q^2),A_{0,1,2,V}(Q^2)$ are the weak transition form factors. The form factors $F_{\pm}(Q^2)$ are related to the commonly used Bauer-Stech-Wirbel form factors via
\begin{eqnarray}
    F_+(Q^2)&=&F_1(Q^2),\\
    F_-(Q^2)&=&\frac{m_B^2-m_{D^\ast}^2}{Q^2}\left(F_1(Q^2)-F_0(Q^2)\right).
\end{eqnarray}
\par With the above hadronic matrix elements, the weak decay amplitude can be expressed as:
\begin{eqnarray}
\mathcal{A}(B^+\rightarrow D_s^{\ast+}\bar{D}^0)&=&\mathrm{i}\sqrt{2}G_FV_{cb}V_{cs}^\ast a_1 \lambda_{D_s^\ast}F_+^{B\rightarrow D}(-m_{D_s^\ast}^2)\epsilon_{D_s^{\ast}}^\ast\cdot p_{B^+}\\
\mathcal{A}(B^+\rightarrow D_{s1}^+\bar{D}^{\ast0})&=&-\frac{G_F}{\sqrt{2}}\frac{V_{cb}V_{cs}^\ast a_1 \lambda_{D_{s1}}}{m_B+m_{D^\ast}}\left((m_B+m_{D^\ast})^2(\epsilon_{D^\ast}\cdot\epsilon_{D_{s1}})A_1(-m_{D_{s1}}^2)-2(\epsilon_{D^\ast}\cdot p_B)(\epsilon_{D_{s1}}\cdot p_B)A_2(-m_{D_{s1}}^2)\right)
\end{eqnarray}
The amplitude for $B^+\rightarrow D_s^{\ast+}\bar{D}^0\rightarrow D^+X_0(2900)$ process can be written as:
\begin{equation}\label{Eq:AmpB-DsD0-DX}
\begin{split}
&\mathcal{A}(B^+\rightarrow D_s^{\ast+}\bar{D}^0\rightarrow D^+X_0(2900))\\
=&\mathrm{i}\sqrt{2}G_FV_{cb}V_{cs}^\ast a_1m_{X_0}\lambda_{D_{s}^\ast}F_+^{B\rightarrow D}(-m_{D_s^\ast}^2)\int_{-1}^1\frac{|p_{D_s^\ast}|\mathrm{dcos}\theta\mathrm{d}\phi}{32\pi^2m_B}\frac{\mathrm{i}g_{D_s^\ast DK}(-t)g_{DKX_0}(-t)}{t-m_K^2}H_1,
\end{split}
\end{equation}
where
\begin{equation}
H_1=\left(\frac{(p_{D^0}\cdot p_{D_s^\ast})((p_{D^+}\cdot p_{D_s^\ast})}{m_{D_s^\ast}^2}-(p_{D^+}\cdot p_{D^0})\right)(p_{D_s^\ast}\cdot p_{D^0}-p_{D^+}\cdot p_{D^0}).
\end{equation}
The amplitude for $B^+\rightarrow D_{s1}\bar{D}^\ast\rightarrow D^+X_0(2900)$ process can be written as:
\begin{equation}\label{Eq:AmpB-Ds1D-DX}
\begin{split}
&\mathcal{A}(B^+\rightarrow D_{s1}\bar{D}^\ast\rightarrow D^+X_0(2900))\\
=&-\mathrm{i}\frac{G_F}{\sqrt{2}}V_{cb}V_{cs}^\ast a_1\frac{\mathrm{i}\lambda_{D_{s1}}}{(m_B+m_D^\ast)m_{D^\ast}^2m_{D_{s1}}^2m_{K^\ast}^2}\int_{-1}^1\frac{|p_{D^\ast}|\mathrm{dcos}\theta\mathrm{d}\phi}{32\pi^2m_B}\frac{g_{D_{s1}DK^\ast}(-t)g_{D^\ast K^\ast X_0}(-t)}{t-m_{K^\ast}^2}\Big(H_2 A_1(-m_{D_{s1}}^2)+2 H_3 A_2(-m_{D_{s1}}^2)\Big),
\end{split}
\end{equation}
where 
\begin{equation}
\begin{split}
H_2=&m_{D^\ast}^2m_{D_{s1}}^2(m_B+m_{D^\ast})^2(m_D^2-2m_{K^\ast}^2)+m_{K^{\ast}}^2(p_{D^\ast}\cdot p_{D_{s1}})^2+m_{D^\ast}^2(p_{D}\cdot p_{D_{s1}})^2\\
&+(p_{D^\ast}\cdot p_{D}-p_{D^\ast}\cdot p_{D_{s1}})\Big(m_{D_{s1}}^2p_{D^\ast}\cdot p_{D}-(p_{D^\ast}\cdot p_{D_{s1}})\cdot(p_{D}\cdot p_{D_{s1}})\Big),\\
H_3=&m_{K^\ast}^2p_{D^\ast}\cdot p_{D_{s1}}\Big((p_{D^\ast}\cdot p_{D_{s1}})^2-m_{D^\ast}^2m_{D_{s1}}^2\Big)+\Big(m_{D^\ast}^2 p_{D^\ast}\cdot p_D-(p_{D_{s1}}\cdot p_D)(p_{D^\ast}\cdot p_{D_{s1}})\Big)\\
&\Big(m_{D^\ast}^2(m_{D_{s1}}^2-p_{D_{s1}}\cdot p_{D})+p_{D^\ast}\cdot p_{D_{s1}}(p_{D^\ast}\cdot p_D-p_{D^\ast}\cdot p_{D_{s1}})\Big).
\end{split}
\end{equation}
Similarly, the amplitude for the other processes shown in Fig.~\ref{Fig:BDXhadronlevel} can be written as:
\allowdisplaybreaks{
\begin{eqnarray}
\nonumber \mathcal{A}(B^+\rightarrow D_{s}\bar{D}^\ast\rightarrow D^+X_0(2900))&=&-\mathrm{i}\frac{G_F}{\sqrt{2}}V_{cb}V_{cs}^\ast a_1\frac{\mathrm{i}\lambda_{D_{s1}}}{m_{D^\ast}^2m_{K^\ast}^2}\Big((m_B+m_{D^\ast})A_1(-m_{D_s}^2)-2(m_B-m_{D^\ast})A_2(-m_{D_s}^2) \\
& &-2m_{D^\ast}(A_3(-m_{D_s}^2)-A_0(-m_{D_s}^2))\Big)\int_{-1}^1\frac{|p_{D_{s}}|\mathrm{dcos}\theta\mathrm{d}\phi}{32\pi^2m_B}\frac{g_{D_{s} DK^\ast}(-t)g_{D^\ast K^\ast X_0}(-t)}{t-m_{K^\ast}^2}H_4,\\
\nonumber \mathcal{A}(B^+\rightarrow D_{s0}^\ast\bar{D}\rightarrow D^+X_0(2900))&=&\mathrm{i}\frac{G_F}{\sqrt{2}}V_{cb}V_{cs}^\ast a_1m_{X_0}\lambda_{D_{s}^\ast}\left(F_+^{B\rightarrow D}(-m_{D_{s0}^\ast}^2)(m_B^2-m_D^2)+F_-^{B\rightarrow D}(-m_{D_{s0}^\ast}^2)\right)\\
& &\int_{-1}^1\frac{|p_{D_{s0}^\ast}|\mathrm{dcos}\theta\mathrm{d}\phi}{32\pi^2m_B}\frac{\mathrm{i}g_{D_{s0}^\ast DK}(-t)g_{DKX_0}(-t)}{t-m_K^2}\left(p_{D^+}\cdot p_{D^\ast_{s0}}-m_D^2\right)(p_{D_{s0}^\ast}\cdot p_{D^0}-p_{D^+}\cdot p_{D^0}),\\
\nonumber\mathcal{A}(B^+\rightarrow D_{s}^\ast\bar{D}_1\rightarrow D^+X_0(2900))&=&\mathrm{i}\frac{G_F}{\sqrt{2}}V_{cb}V_{cs}^\ast a_1\frac{\mathrm{i}\lambda_{D_{s}^\ast}}{(m_B+m_{D1})m_{D_1}^2m_{D_{s}^\ast}^2m_{K_1}^2}\\
& &\int_{-1}^1\frac{|p_{D_1}|\mathrm{dcos}\theta\mathrm{d}\phi}{32\pi^2m_B}\frac{g_{D_{s}^\ast DK_1}(-t)g_{D_1 K_1 X_0}(-t)}{t-m_{K_1}^2}\Big(H_5 A_1(-m_{D_{s}^\ast}^2)+2 H_6 A_2(-m_{D_{s}^\ast}^2)\Big),\\
\nonumber\mathcal{A}(B^+\rightarrow D_{s0}^\ast\bar{D}_1\rightarrow D^+X_0(2900))&=&\mathrm{i}\frac{G_F}{\sqrt{2}}V_{cb}V_{cs}^\ast a_1\frac{\mathrm{i}\lambda_{D_{s}^\ast}}{m_{D_1}^2m_{K_1}^2}\Big((m_B+m_{D_1})A_1(-m_{D_{s0}^\ast}^2)-2(m_B-m_{D_1})A_2(-m_{D_{s0}^\ast}^2) \\
& &-2m_{D_1}(A_3(-m_{D_{s0}^\ast}^2)-A_0(-m_{D_{s0}^\ast}^2))\Big)\int_{-1}^1\frac{|p_{D_{s0}^\ast}|\mathrm{dcos}\theta\mathrm{d}\phi}{32\pi^2m_B}\frac{g_{D_{s0}^\ast DK_1}(-t)g_{D_1K_1 X_0}(-t)}{t-m_{K_1}^2}H_7,\\
\nonumber \mathcal{A}(B^+\rightarrow D_{s}\bar{D}^\ast_0\rightarrow D^+X_0(2900))&=&-\mathrm{i}\frac{G_F}{\sqrt{2}}V_{cb}V_{cs}^\ast a_1m_{X_0}\lambda_{D_{s1}}\left(F_+^{B\rightarrow D_0^\ast}(-m_{D_{s}^\ast}^2)(m_B^2-m_D^2)+F_-^{B\rightarrow D_0^\ast}(-m_{D_{s}^\ast}^2)\right)\\
& &\int_{-1}^1\frac{|p_{D_{s}}|\mathrm{dcos}\theta\mathrm{d}\phi}{32\pi^2m_B}\frac{\mathrm{i}g_{D_{s} DK_0^\ast}(-t)g_{D_0^\ast K_0^\ast X_0}(-t)}{t-m_{K_0^\ast}^2}\left(p_{D^+}\cdot p_{D_{s}}\right)(p_{D_s}\cdot p_{D^\ast}-p_{D^+}\cdot p_{D^\ast}),\\
\mathcal{A}(B^+\rightarrow D_{s1}\bar{D}^\ast_0\rightarrow D^+X_0(2900))&=&-\mathrm{i}\sqrt{2}G_FV_{cb}V_{cs}^\ast a_1m_{X_0}\lambda_{D_{s1}}F_+^{B\rightarrow D}(-m_{D_s^\ast}^2)\int_{-1}^1\frac{|p_{D_s^\ast}|\mathrm{dcos}\theta\mathrm{d}\phi}{32\pi^2m_B}\frac{\mathrm{i}g_{D_s^\ast DK}(-t)g_{DKX_0}(-t)}{t-m_K^2}H_8,\label{Eq:AmpB-Ds1D0-DX}
\end{eqnarray}
}
where
\allowdisplaybreaks{
\begin{eqnarray}
\nonumber H_4&=&(m_D^2-m_{D^\ast}^2+m_{K^\ast}^2)\Big(m_{D_s}^2m_{D^\ast}^2-(p_{D_s}\cdot p_{\bar{D}^{\ast0}})^2\Big)\\
& &+(m_D^2-m_{D^\ast}^2-m_{K^\ast}^2)\Big((p_{D_s}\cdot p_{\bar{D}^{\ast0}})(p_{D^+}\cdot p_{\bar{D}^{\ast0}})-m_{D^\ast}^2(p_{D_s}\cdot p_{D^+})\Big),\\
\nonumber H_5&=&m_{D_1}^2m_{D_{s}^\ast}^2(m_B+m_{D_1})^2(m_D^2-2m_{K_1}^2)+m_{K_1}^2(p_{D_1}\cdot p_{D_{s}^\ast})^2+m_{D_1}^2(p_{D}\cdot p_{D_{s}^\ast})^2\\
& &+(p_{D_1}\cdot p_{D}-p_{D_1}\cdot p_{D_{s}^\ast})\Big(m_{D_{s}^\ast}^2p_{D_1}\cdot p_{D}-(p_{D_1}\cdot p_{D_{s}^\ast})\cdot(p_{D}\cdot p_{D_{s}^\ast})\Big),\\
\nonumber H_6&=&m_{K_1}^2p_{D_1}\cdot p_{D_{s}^\ast}\Big((p_{D_1}\cdot p_{D_{s}^\ast})^2-m_{D_1}^2m_{D_{s}^\ast}^2\Big)+\Big(m_{D_1}^2 p_{D_1}\cdot p_D-(p_{D_{s}^\ast}\cdot p_D)(p_{D_1}\cdot p_{D_{s}^\ast})\Big)\\
& &\Big(m_{D_1}^2(m_{D_{s}^\ast}^2-p_{D_{s}^\ast}\cdot p_{D})+p_{D_1}\cdot p_{D_{s}^\ast}(p_{D_1}\cdot p_D-p_{D_1}\cdot p_{D_{s}^\ast})\Big),\\
\nonumber H_7&=&m_D^2\Big(m_{D_{s0}}^2m_{D_1}^2-(p_{D_{s0}}\cdot p_{D_1})^2\Big)+(p_{D_{s0}}\cdot p_{D_1})(p_{D^+}\cdot p_{D_1})(m_D^2-m_{K_1}^2-p_{D_{s0}}\cdot p_{D^+})\\
& &+(p_{D_{s0}}\cdot p_{D^+})\Big(m_{D_1}^2(p_{D_{s0}}\cdot p_{D^+}-m_D^2-m_{D_{s0}}^2+m_{K_1}^2)+(p_{D_{s0}}\cdot p_{D_1})^2\Big),\\
H_8&=&\left(\frac{(p_{D^\ast}\cdot p_{D_{s1}})((p_{D^+}\cdot p_{D_{s1}})}{m_{D_{s1}}^2}-(p_{D^+}\cdot p_{D^\ast})\right)(p_{D_{s1}}\cdot p_{D^\ast}-p_{D^+}\cdot p_{D^\ast}).
\end{eqnarray}
}
\par It should be noted that in Ref.\cite{Chen:2020eyu}, the decay amplitude in Eq.~\eqref{Eq:AmpB-DsD0-DX} contains the form factor $F(t,m)=(\Lambda^2-m_K^2)/(\Lambda^2-t)$ for each strong vertices, which is introduced to compensate the off-shell effect of the exchanged particle at the vertices~\cite{Gortchakov:1995im}. In Ref.~\cite{Cheng:2003sm}, the authors claim that the monopole behavior of the form factor is preferred as it is consistent with the QCD sum rule expectation. The cutoff $\Lambda$ is usually chosen as
\begin{equation}
\Lambda=m_K+\eta \Lambda_{\mathrm{QCD}},
\end{equation}
and the amplitude would be very sensitive to the value of $\eta$. However, we will see with QCD sum rule approach in the next section, the form factor does not always behave as a monopole model. With the above consideration, we shall replace the $g_{ABC}$ with $F(t,m)$ given in the Ref.~\cite{Chen:2020eyu} by $g_{ABC}(Q^2)$ for a $ABC$ strong vertex shown in Eq.~\eqref{Eq:AmpB-DsD0-DX} and ~\eqref{Eq:AmpB-Ds1D-DX}. In the next section, we shall obtain the coupling constant in Eq.~\eqref{Eq:AmpB-DsD0-DX} and ~\eqref{Eq:AmpB-Ds1D-DX} with QCD sum rule approach and show that the coupling constant $g_{ABC}(Q^2)$ does not always behave as a monopole model.
\section{Three-point QCD sum rule}\label{Sec.QCDSR}
\par Over past several decades, the method of QCD sum rules has been proven to be very powerful to study hadron properties~\cite{Reinders:1984sr,Shifman:1978bx,Colangelo:2000dp,Narison:2002woh}. In this section, we shall study the three-point correlation function of several two-body strong decay process $M\rightarrow X+Y$. For the strong decay process $M\rightarrow X+Y$, the corresponding correlator is written as
\begin{equation}\label{Eq:StrongDecayCF}
\Pi(p,p^{'},q)=\int\mathrm{d}^4x\mathrm{d}^4y\;\mathrm{e}^{\mathrm{i}p^{'}\cdot x}\mathrm{e}^{\mathrm{i}q\cdot y}\langle0|T\{J_{X}(x)J_Y(y)J^\dagger_M(0)\}|0\rangle,
\end{equation}
where $J_{M(X,Y)}$ is the interpolating current for the initial(final) state. In this section, we shall consider the $\bar{D}KX_0$, $\bar{D}^\ast K^\ast X_0$, $\bar{D}_1K_1X_0$, $\bar{D}_0^\ast K_0^\ast X_0$ strong decay vertices with $K(K^\ast,K_1,K_0^\ast)$ off shell and $D_{s}^\ast DK$, $D_{s1}DK^\ast$, $D_{s0}^\ast DK$, $D_{s}DK^\ast$, $D_{s}^\ast DK_1$, $D_{s0}DK_1^\ast$, $D_{s}^\ast DK_0^\ast$, $D_{s1}DK_0^\ast$ strong decay vertices with $K_(K^\ast,K_1,K_0^\ast)$ off shell. Some of these strong decay vertices are studied by some previous QCD sum rule analysis. We use the following interpolating currents for $X_0(2900)$ by considering it as a $\bar{D}^\ast K^\ast$ molecule, scalar diquark-scalar antidiquark compact tetraquark and axial-vector diquark-axial-vector antidiquark compact tetraquark:
\begin{eqnarray}
J_{X_0(\mathrm{mol})}&=&\bar{c}_a\gamma_\mu d_a \bar{s}_b\gamma_\mu u_b,\\
J_{X_0(\mathrm{S-S})}&=&\epsilon_{abc}\epsilon_{efc}(u_a^T\mathcal{C}\gamma_5d_b)(\bar{c}_e\gamma_5\mathcal{C}\bar{s}_f^T),\\
J_{X_0(\mathrm{A-A})}&=&\epsilon_{abc}\epsilon_{efc}(u_a^T\mathcal{C}\gamma_\mu d_b)(\bar{c}_e\gamma_\mu\mathcal{C}\bar{s}_f^T),
\end{eqnarray}
where $a\cdots f$ denote the color index and $u,d,s,c$ denote the up, down, strange, charm quark field, respectively. These current can couple to the $X_0(2900)$ state via
\begin{eqnarray}
\langle0|J_{X_0(\mathrm{mol})}|X_0(2900)\rangle&=&\lambda_{X_0(\mathrm{mol})},\\
\langle0|J_{X_0(\mathrm{S-S})}|X_0(2900)\rangle&=&\lambda_{X_0(\mathrm{S-S})},\\
\langle0|J_{X_0(\mathrm{A-A})}|X_0(2900)\rangle&=&\lambda_{X_0(\mathrm{A-A})},
\end{eqnarray}
in which the value of the coupling constant $\lambda_{X_0}$ are determined from the two-point mass sum rules established in Ref.~\cite{Agaev:2020nrc,Zhang:2020oze,Wang:2020xyc} :
\begin{equation}
\begin{split}
\lambda_{X_0(\mathrm{mol})}/m_{X_0}&=(3.0\pm0.7)\times10^{-3}\mathrm{GeV}^5,\\
\lambda_{X_0(\mathrm{S-S})}/m_{X_0}&=(1.2\pm0.2)\times10^{-2}\mathrm{GeV}^5,\\
\lambda_{X_0(\mathrm{A-A})}/m_{X_0}&=(1.6\pm0.3)\times10^{-2}\mathrm{GeV}^5.\\
\end{split}
\end{equation}
The other hadronic parameters in this work are listed in Tab.~\ref{Tab:Hadronicinput}.
\begin{table}[htbp]
\caption{The values of the hadronic parameters $m_H$ and $f_H$ in the work taken from Ref.~\cite{ParticleDataGroup:2022pth,Gelhausen:2013wia,Gubernari:2023rfu}.}\label{Tab:Hadronicinput}\renewcommand\arraystretch{1.6} 
\setlength{\tabcolsep}{0.5 em}{ 
\centering
\begin{tabular}{c c c|c c c}
  \hline
 \hline
Hadron(H) & Mass $m_H$[GeV] & Decay constant $f_H$[GeV] & Hadron(H) &  Mass $m_H$[GeV] & Decay constant $f_H$[GeV] \\
\hline
$D$ & 1.87 & $0.20\pm0.02$ & $D^{\ast}$ & 2.01 & $0.24\pm0.04$ \\
$D_s$ & 1.97 & $0.24\pm0.04$ & $D_s^{\ast}$ & 2.11 & $0.29\pm0.05$ \\
$K$ & 0.49 & $0.16\pm0.02$ & $K^{\ast}$ & 0.89 & $0.22\pm0.01$ \\
$D_{1}$ & 2.42 & $0.27\pm0.01$ & $D_{0}^{\ast}$ & 2.34 & $0.10\pm0.01$ \\
$D_{s1}$ & 2.46 & $0.23\pm0.02$ & $D_{s0}^{\ast}$ & 2.32 & $0.18\pm0.03$ \\
$K_{1}$ & 1.25 & $0.25\pm0.01$ & $K_0^{\ast}$ & 0.85 & $0.19\pm0.01$ \\
 \hline
  \hline
\end{tabular}}
\end{table}
\subsection{Coupling constant of $X_0(2900)$}
\par
\begin{figure}[htbp]
\centering
\includegraphics[width=\textwidth]{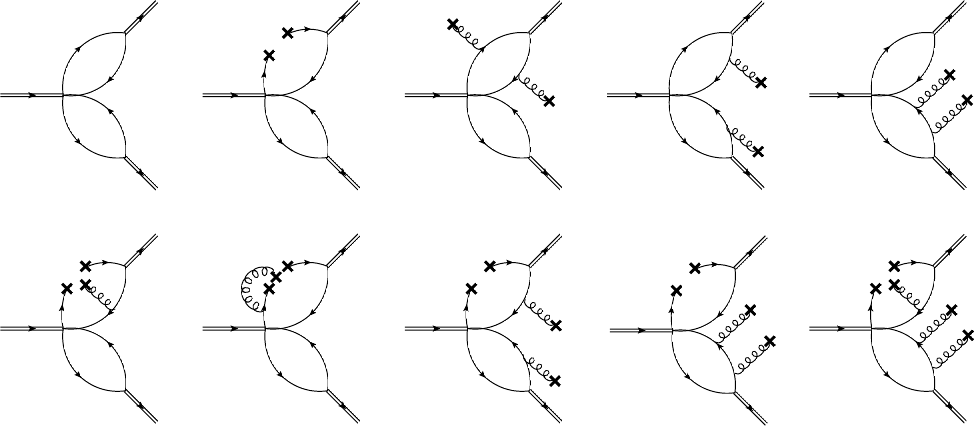}
\caption{The Feynman diagrams for OPE series of the correlation function in \eqref{Eq:gDKX}. The double solid line denotes the in(out)-coming hadron, while the single solid line denotes the quark propagator. Diagrams related by symmetry are not shown.}
\label{Fig:FeynDKX}
\end{figure}

 In this subsection, we study the strong vertex $\bar{D}KX_0$ and $\bar{D}^\ast K^\ast X_0$ with $\bar{D}^\ast K^\ast$ molecule interpretation for $X_0(2900)$ state as an example. The interpolating currents for $\bar{D}^{(\ast)}$ and $K^{\ast0}$ mesons are 
\begin{equation}
\begin{split}
J_{\bar{D}}&=\mathrm{i}\bar{c}_a\gamma_5u_a,\\
J_{\bar{D}^\ast\mu}&=\bar{c}_a\gamma_\mu u_a,\\
J_{K^0}&=\mathrm{i}\bar{s}_a\gamma_5d_a,\\
J_{K^{\ast0}\mu}&=\bar{s}_a\gamma_\mu d_a.
\end{split}
\end{equation}
They can couple to the corresponding mesons via the following relations
\begin{equation}
\begin{split}
\langle0|J_{\bar{D}}|\bar{D}\rangle&=f_{\bar{D}}\frac{m_{\bar{D}}^2}{m_c}\equiv\lambda_{\bar{D}},\\
\langle0|J_{K^0}|K^0\rangle&=f_K\frac{m_K^2}{m_s}\equiv\lambda_{K},\\
\langle0|J_{\bar{D}^\ast\mu}|\bar{D}^\ast\rangle&=m_{\bar{D}^\ast}f_{\bar{D}^\ast}\epsilon_\mu\equiv\lambda_{\bar{D}^\ast}\epsilon_\mu,\\
\langle0|J_{K^{\ast0}\mu}|K^{\ast0}\rangle&=m_{K^\ast}f_{K^\ast}\epsilon_\mu\equiv\lambda_{K^\ast}\epsilon_\mu.
\end{split}
\end{equation}
Here $f_{\bar{D}}$, $f_{K}$, $f_{\bar{D}^*}$, $f_{K^*}$ are the decay constants of the $\bar{D}$, $K$, $\bar{D}^*$, $K^*$ mesons. $\epsilon_\mu$ is the polarization vector.
The coupling constant $g_{\bar{D}KX_0}$ and $g_{\bar{D}^\ast K^\ast X_0}$ are defined via the effective Lagrangian~\cite{tHooft:2008rus,Xiao:2020ltm}
\begin{eqnarray}
\mathcal{L}_{\bar{D}KX_0}&=&g_{\bar{D}KX_0}m_{X_0}X_0\partial^\mu\bar{D}\partial_\mu K,\\
\mathcal{L}_{\bar{D}^\ast K^\ast X_0}&=&g_{\bar{D}^\ast K^\ast X_0}X_0\bar{D}^{\ast\mu} K^\ast_\mu,
\end{eqnarray}
thus the transition matrix element can be obtained as
\begin{eqnarray}\label{Eq:MatrixelementofDKX}
\langle\bar{D}(p)K^0(q)|X_0(p^{'})\rangle&=&g_{\bar{D}KX_0}m_{X_0}p\cdot q,\\
\langle\bar{D}^\ast(p)K^{\ast0}(q)|X_0(p^{'})\rangle&=&g_{\bar{D}^\ast K^\ast X_0} \epsilon(p)\cdot\epsilon(q).
\end{eqnarray}
With the above coupling relations and transition matrix element, we can obtain the three-point correlation function Eq.\eqref{Eq:StrongDecayCF} for $\bar{D}K\rightarrow X_0$ process on the phenomenological side 
\begin{equation}
\begin{split}
\Pi(p,p^{'},q)=&\int\mathrm{d}^4x\mathrm{d}^4y\;\mathrm{e}^{\mathrm{i}p^{'}\cdot x}\mathrm{e}^{-\mathrm{i}q\cdot y}\langle0|T\{J_{X_0}(x)J^\dagger_{K^0}(y)J^\dagger_{\bar{D}}(0)\}|0\rangle\\
=&\frac{\lambda_{X_0}\lambda_{\bar{D}}\lambda_{K}g_{\bar{D}KX_0}m_{X_0}(p\cdot q)}{(p^{'2}-m_{X_0}^2)(p^2-m_D^2)(q^2-m_K^2)}+\cdots.
\end{split}
\end{equation}
On the OPE side, we can evaluate the correlation function with standard QCD sum rules approach. To establish a sum rule for the coupling constant, we will pick out the $1/q^2$ terms around the pole $q^2\sim0$ with the structure $p\cdot q$ in the OPE series and then match both sides of the sum rule. To apply sum rules appropriately, we shall calculated at $Q^2$ far away from the on-shell mass $-m_K^2$ to ensure the approximation $p^2=p^{'2}=P^2$ valid. After performing the Borel transform $P^2\rightarrow M_B^2$ on both phenomenological and OPE sides, we obtain
\begin{equation}\label{Eq:gDKX}
g_{\bar{D}KX_0}(s_0,M_B^2)=\frac{1}{\lambda_{X_0}\lambda_{\bar{D}}\lambda_{K}m_{X_0}}\frac{m_{X_0}^2-m_D^2}{\mathrm{e}^{-m_D^2/M_B^2}-\mathrm{e}^{-m_{X_0}^2/M_B^2}}\left(\frac{Q^2+m_K^2}{Q^2}\right)\left(\int_{s_<}^{s_0}\mathrm{d}s\;\rho(s)\mathrm{e}^{-s/M_B^2}+R(M_B^2)\mathrm{e}^{-m_c^2/M_B^2}\right),
\end{equation}
where the continuum threshold $s_0=10\;\mathrm{GeV}^2$ is taken from the two-point mass sum rules in Ref.~\cite{Chen:2020aos}. The spectrum function $\rho(s)$ and $R(M_B^2)$ are given in Appendix A, and the Feynman diagram we considered are shown in Fig.~\ref{Fig:FeynDKX}.
\par We use the standard values of various QCD condensates as $\langle \bar{q}q\rangle(1\mathrm{GeV})=-(0.24\pm0.03)^3\;\mathrm{GeV}^3$, $\langle \bar{q}g_s\sigma\cdot Gq\rangle(1\mathrm{GeV})=-M_0^2\langle \bar{q}q\rangle$, $M_0^2=(0.8\pm0.2)\;\mathrm{GeV}^2$, $\langle \bar{s}s\rangle/\langle \bar{q}q\rangle=0.8\pm0.1$, $\langle g_s^2GG\rangle(1\mathrm{GeV})=(0.48\pm0.14)\;\mathrm{GeV}^4$ at the energy scale $\mu=1$GeV~\cite{Narison:1989aq,Jamin:2001zr,Jamin:1998ra,Ioffe:1981kw,Chung:1984gr,Dosch:1988vv,Khodjamirian:2011ub,Francis:2018jyb} and $m_s(2\;\mathrm{GeV})=95^{+9}_{-3}\;\mathrm{MeV}$, $m_c(m_c)=1.27^{+0.03}_{-0.04}\;\mathrm{GeV}$, $m_b(m_b)=4.18_{-0.03}^{+0.04}\;\mathrm{GeV}$ from the Particle Data Group\cite{ParticleDataGroup:2022pth}. These values are widely used in the previous QCD sum rules for traditional baryon states, meson states, multiquark states, etc. , and provide experimentally consistent results~\cite{Narison:1989aq,Reinders:1984sr,Albuquerque:2018jkn,Chen:2016qju}. We also take into account the energy-scale dependence of the above parameters from the renormalization group equation
\allowdisplaybreaks{
\begin{eqnarray}\label{inputparameter}
\nonumber&&m_s(\mu)=m_s(2\mathrm{GeV})\left[\frac{\alpha_s(\mu)}{\alpha_s(2\mathrm{GeV})}\right]^{\frac{12}{33-2n_f}},\\
\nonumber&&m_c(\mu)=m_c(m_c)\left[\frac{\alpha_s(\mu)}{\alpha_s(m_c)}\right]^{\frac{12}{33-2n_f}},\\
\nonumber&&m_b(m_b)=m_b(m_b)\left[\frac{\alpha_s(\mu)}{\alpha_s(m_b)}\right]^{\frac{12}{33-2n_f}},\\
\nonumber&&\langle \bar{q}q\rangle(\mu)=\langle \bar{q}q\rangle(1\mathrm{GeV})\left[\frac{\alpha_s(1\mathrm{GeV})}{\alpha_s(\mu)}\right]^{\frac{12}{33-2n_f}},\\
  &&\langle \bar{s}s\rangle(\mu)=\langle \bar{s}s\rangle(1\mathrm{GeV})\left[\frac{\alpha_s(1\mathrm{GeV})}{\alpha_s(\mu)}\right]^{\frac{12}{33-2n_f}},\\
\nonumber&&\langle \bar{q}g_s\sigma\cdot Gq\rangle(\mu)=\langle \bar{q}g_s\sigma\cdot Gq\rangle(1\mathrm{GeV})\left[\frac{\alpha_s(1\mathrm{GeV})}{\alpha_s(\mu)}\right]^{\frac{2}{33-2n_f}},\\
\nonumber&&\langle \bar{s}g_s\sigma\cdot Gs\rangle(\mu)=\langle \bar{s}g_s\sigma\cdot Gs\rangle(1\mathrm{GeV})\left[\frac{\alpha_s(1\mathrm{GeV})}{\alpha_s(\mu)}\right]^{\frac{2}{33-2n_f}},\\  
\nonumber&&\alpha_s(\mu)=\frac{1}{b_0t}\left[1-\frac{b_1}{b_0}\frac{\mathrm{log}t}{t}+\frac{b_1^2(\mathrm{log}^2t-\mathrm{log}t-1)+b_0b_2}{b_0^4t^2}\right],
\end{eqnarray}
}
where $t=\mathrm{log}\frac{\mu^2}{\Lambda^2}$, $b_0=\frac{33-2n_f}{12\pi}$, $b_1=\frac{153-19n_f}{24\pi^2}$, $b_2=\frac{2857-\frac{5033}{9}n_f+\frac{325}{27}n_f^2}{128\pi^3}$, $\Lambda=$210 MeV, 292 MeV and 332 MeV for the flavors $n_f=$5, 4 and 3, respectively. In this work, we evolve all the input parameters to the energy scale $\mu=m_c$ for our sum rule analysis. The parameters for the $D^{(\ast)}$ and $K^{(\ast)}$ mesons are adopted in Tab.~\ref{Tab:Hadronicinput}.
\begin{figure}[htbp]
\centering
\includegraphics[width=8cm]{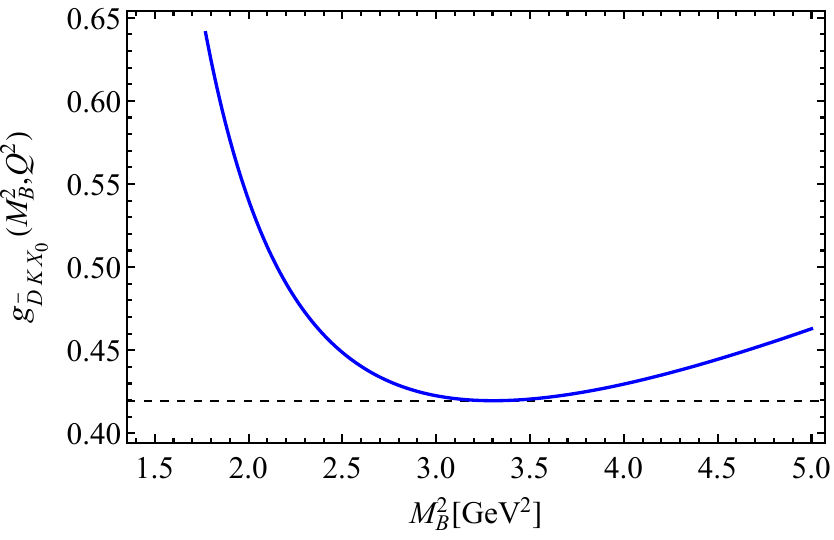}\quad
\includegraphics[width=8cm]{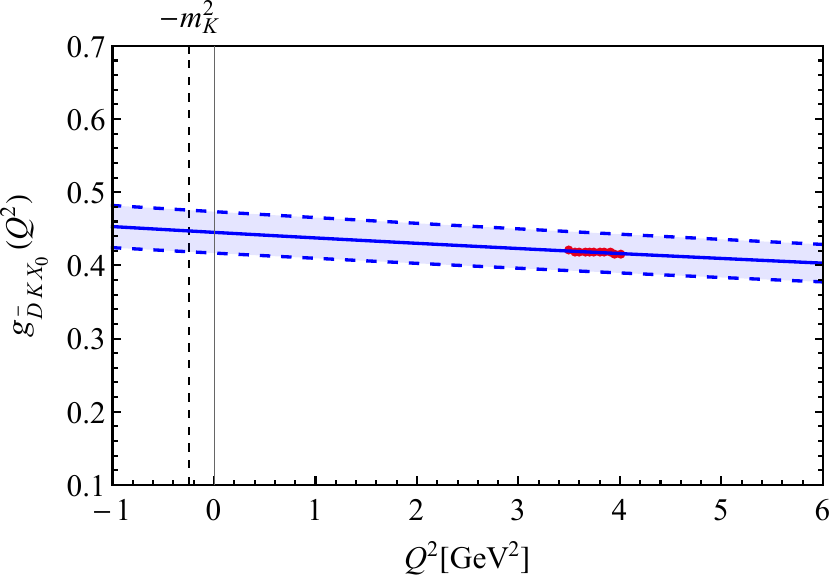}\\
\caption{The dependence of the coupling constant $g_{\bar{D}KX_0}$ on the Borel mass $M_B^2$ (left panel) and transfer momentum $Q^2$ (right panel). On the left panel, the transfer momentum is set to be $Q^2=m_D^2\sim 3.5\;\mathrm{GeV}^2$. On the right panel, the red dots denote the value from Eq.\eqref{Eq:gDKX} with $s_0=10\;\mathrm{GeV}^2$ and $M_B^2=3.30\;\mathrm{GeV}^2$. The blue solid line is the monopolar fitting curve. The two dashed blue lines denote the upper and lower boundary of the uncertainty from various condensates, quark masses and hadronic parameters.}
\label{Fig:gDKXMB&Q2}
\end{figure}
\par In the left panel of Fig.\ref{Fig:gDKXMB&Q2}, we show the variation of the coupling constant $g_{\bar{D}KX_0}(Q^2)$ with the Borel mass $M_B^2$ at $Q^2=m_D^2\sim 3.5\;\mathrm{GeV}^2$. Such a momentum point is chosen far away from $m_K^2$ so that it can be safely ignored and the OPE series is valid in this region. We find that the coupling constant $g_{\bar{D}KX_0}(Q^2)$ has a minimum value at $M_B^2\sim 3.30\;\mathrm{GeV}^2$, around which it has minimal dependence on the non-physical parameter $M_B^2$. To extrapolate the coupling constant from the valid QCD sum rule working region to the physical pole $Q^2=-m_K^2$, we fit the sum rule result for $s_0=10\;\mathrm{GeV}^2$ and $M_B^2=3.30\;\mathrm{GeV}^2$ with monopolar model 
\begin{equation}
g_{\bar{D}KX_0}(Q^2)=\frac{a}{b+Q^2}.
\end{equation}
The fitting curve is shown in the right panel of Fig.\ref{Fig:gDKXMB&Q2}, and the result is as follow:
\begin{equation}
g_{\bar{D}KX_0}(Q^2)=\frac{(25.56\pm 1.63)\;\mathrm{GeV}^2}{Q^2+(57.43\pm0.00)\;\mathrm{GeV}^2}.
\end{equation}
\par As for $\bar{D}^\ast K^\ast\rightarrow X_0$ process
\begin{equation}
\begin{split}
\Pi(p,p^{'},q)=&\int\mathrm{d}^4x\mathrm{d}^4y\;\mathrm{e}^{\mathrm{i}p^{'}\cdot x}\mathrm{e}^{-\mathrm{i}q\cdot y}\langle0|T\{J_{X_0}(x)J^\dagger_{K^{\ast0}\mu}(y)J^\dagger_{\bar{D}^\ast\mu}(0)\}|0\rangle\\
=&\frac{\lambda_{X_0}\lambda_{\bar{D}^\ast}\lambda_{K^\ast}g_{\bar{D}^\ast K^\ast X_0}}{(p^{'2}-m_{X_0}^2)(p^2-m_{D^\ast}^2)(q^2-m_{K^\ast}^2)}\left(3+\frac{Q^2}{m_{K^\ast}^2}+\frac{(Q^2-m_{D^\ast}^2+m_{X_0}^2)^2}{4m_{D^\ast}^2m_{K^\ast}^2}\right)+\cdots.
\end{split}
\end{equation}
After performing the Borel transform $P^2\rightarrow M_B^2$ on both phenomenological and OPE sides, we obtain
\begin{equation}\label{Eq:gDsKsX}
\begin{split}
g_{\bar{D}^\ast K^\ast X_0}(s_0,M_B^2)=&\frac{1}{\lambda_{X_0}\lambda_{\bar{D}^\ast}\lambda_{K^\ast}m_{X_0}}\frac{m_{X_0}^2-m_D^{\ast2}}{\mathrm{e}^{-m_D^{\ast2}/M_B^2}-\mathrm{e}^{-m_{X_0}^2/M_B^2}}\frac{Q^2+m_K^{\ast2}}{Q^2}\\
&\left(\frac{4 m_{D^\ast}^2m_{K^\ast}^2}{12 m_{D^\ast}^2m_{K^\ast}^2+\left((m_{X_0}-m_{D^\ast})^2+Q^2\right)\left((m_{X_0}+m_{D^\ast})^2+Q^2\right)}\right)\left(\int_{s_<}^{s_0}\mathrm{d}s\;\rho(s)\mathrm{e}^{-s/M_B^2}+R(M_B^2)\mathrm{e}^{-m_c^2/M_B^2}\right),
\end{split}
\end{equation}
where the spectral function $\rho(s)$ and $R(M_B^2)$ are given in Appendix A.
\begin{figure}[htbp]
\centering
\includegraphics[width=8cm]{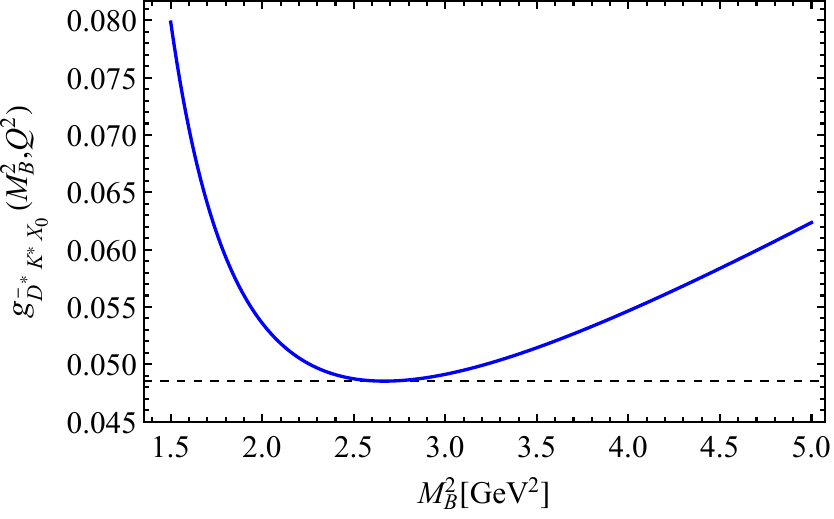}\quad
\includegraphics[width=8cm]{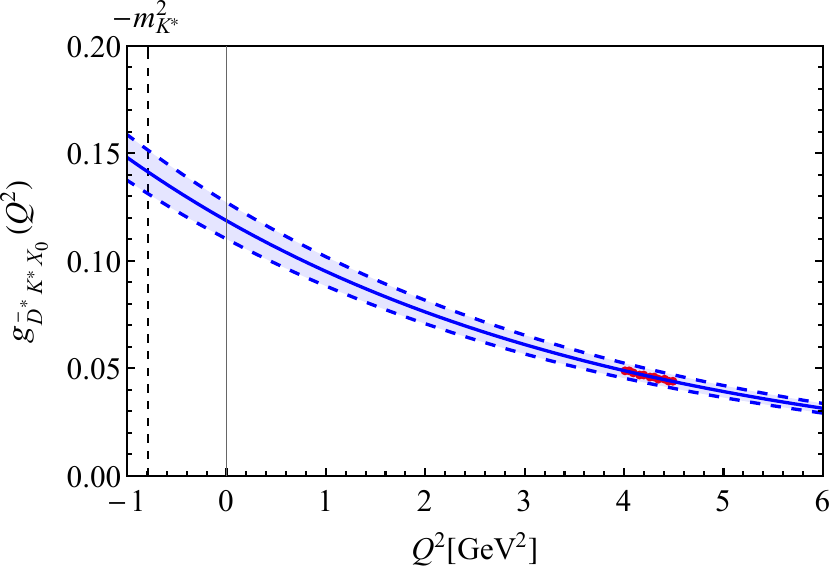}\\
\caption{The dependence of the coupling constant $g_{\bar{D}^\ast K^\ast X_0}$ on the Borel mass $M_B^2$ (left panel) and transfer momentum $Q^2$ (right panel ). On the left panel, the transfer momentum is set to be $Q^2=m_{D^\ast}^2\sim 4\;\mathrm{GeV}^2$. On the right panel, the red dots denote the value from Eq.\eqref{Eq:gDKX} with $s_0=10\;\mathrm{GeV}^2$ and $M_B^2=2.56\;\mathrm{GeV}^2$. The blue solid line is the exponential fitting curve. The two dashed blue lines denote the upper and lower boundary of the uncertainty from various condensates, quark masses and hadronic parameters.}
\label{Fig:gDsKsXMB&Q2}
\end{figure}
\par In the left panel of Fig.\ref{Fig:gDsKsXMB&Q2}, we show the variation of the coupling constant $g_{\bar{D}^\ast K^\ast X_0}(Q^2)$ with the Borel mass $M_B^2$ at $Q^2=m_{D^\ast}^2\sim 4\;\mathrm{GeV}^2$. Such a momentum point is chosen far away from $m_{K^\ast}^2$ so that it can be safely ignored and the OPE series is valid in this region. We find that the coupling constant $g_{\bar{D}^\ast K^\ast X_0}(Q^2)$ has a minimum value at $M_B^2\sim 2.56\;\mathrm{GeV}^2$, around which it has minimal dependence on the non-physical parameter $M_B^2$. We find that the results can be well fitted by the exponential model 
\begin{equation}
g_{\bar{D}^\ast K^\ast X_0}(Q^2)=a\;\mathrm{e}^{-b\;Q^2}.
\end{equation}
The fitting curve is shown in the right panel of Fig.\ref{Fig:gDsKsXMB&Q2}, and the result is as follow:
\begin{equation}
g_{\bar{D}^\ast K^\ast X_0}(Q^2)=(0.12^{+0.008}_{-0.008}\;\mathrm{GeV}^2)\mathrm{e}^{-(0.22\pm0.00)\mathrm{GeV}^{-2}\;Q^2}.
\end{equation}
\par Similarly, we can obtain the coupling constant $g_{D_1K_1X_0}$ and $g_{D_0K_0X_0}$ with interpolating currents
\begin{equation}
\begin{split}
J_{\bar{D}_0^\ast}&=\bar{c}_au_a,\\
J_{\bar{D}_{1\mu}}&=\bar{c}_a\gamma_\mu\gamma_5 u_a,\\
J_{K^0_0}&=\bar{s}_ad_a,\\
J_{K_{1\mu}}&=\bar{s}_a\gamma_\mu\gamma_5 d_a.
\end{split}
\end{equation}
with the following coupling relations
\begin{equation}
\begin{split}
\langle0|J_{\bar{D}_0^\ast}|\bar{D}_0^\ast\rangle&=\lambda_{\bar{D_0^\ast}},\\
\langle0|J_{K_0^\ast}|K_0^\ast\rangle&=\lambda_{K_0^\ast},\\
\langle0|J_{\bar{D}_1^\ast\mu}|\bar{D}_1^\ast\rangle&=\lambda_{\bar{D}_1}\epsilon_\mu,\\
\langle0|J_{K^{0}_1\mu}|K^{0}_1\rangle&=\lambda_{K_1}\epsilon_\mu.
\end{split}
\end{equation}
and the transition matrix
\begin{eqnarray}\label{Eq:MatrixelementofDKX}
\langle\bar{D}_1(p)K^0_1(q)|X_0(p^{'})\rangle&=&g_{\bar{D}_1K_1X_0}\epsilon(p)\cdot\epsilon (q),\\
\langle\bar{D}^\ast_0(p)K^{\ast0}_0(q)|X_0(p^{'})\rangle&=&g_{\bar{D}^\ast_0 K^\ast_0 X_0}m_{X_0}p\cdot q.
\end{eqnarray}
The sum rule for coupling constant $g_{\bar{D}_1K_1X_0}$ and $g_{\bar{D}_0K_0X_0}$ can be obtained as follow:
\begin{eqnarray} 
\nonumber g_{\bar{D}_1 K_1 X_0}(s_0,M_B^2)&=&\frac{1}{\lambda_{X_0}\lambda_{\bar{D}_1}\lambda_{K_1}m_{X_0}}\frac{m_{X_0}^2-m_{D_1}^{2}}{\mathrm{e}^{-m_{D_1}^{2}/M_B^2}-\mathrm{e}^{-m_{X_0}^2/M_B^2}}\frac{Q^2+m_{K_1}^{2}}{Q^2}\\
& &\left(\frac{4 m_{D_1}^2m_{K_1}^2}{12 m_{D_1}^2m_{K_1}^2+\left((m_{X_0}-m_{D_1})^2+Q^2\right)\left((m_{X_0}+m_{D_1})^2+Q^2\right)}\right)\left(\int_{s_<}^{s_0}\mathrm{d}s\;\rho(s)\mathrm{e}^{-s/M_B^2}+R(M_B^2)\mathrm{e}^{-m_c^2/M_B^2}\right),\label{Eq:gD1K1X}\\
g_{\bar{D}_0K_0X_0}(s_0,M_B^2)&=&\frac{1}{\lambda_{X_0}\lambda_{\bar{D}_0^\ast}\lambda_{K_0^\ast}m_{X_0}}\frac{m_{X_0}^2-m_{D_0^\ast}^2}{\mathrm{e}^{-m_{D_0^\ast}^2/M_B^2}-\mathrm{e}^{-m_{X_0}^2/M_B^2}}\left(\frac{Q^2+m_{K_0^\ast}^2}{Q^2}\right)\left(\int_{s_<}^{s_0}\mathrm{d}s\;\rho(s)\mathrm{e}^{-s/M_B^2}+R(M_B^2)\mathrm{e}^{-m_c^2/M_B^2}\right),\label{Eq:gD0K0X}
\end{eqnarray}
The results for coupling constant $g_{\bar{D}_1K_1X_0}$ and $g_{\bar{D}_0K_0X_0}$ are similar as above, as well as other two interpretation of $X_0(2900)$, thus we only list out the results in Tab.~\ref{Tab:FitcoefDKX} and leave out the redundant discussion. The continuum thresholds $s_0$ for the compact structures are taken from Ref.~\cite{Agaev:2022eeh}.
\begin{table*}[!ht]
  \center
  \caption{Fitting coefficients with different models of coupling constant $g_{\bar{D}_{(0,1)}^{(\ast)} K_{(0,1)}^{(\ast)} X_0}$ for different $X_0(2900)$'s structures}
  \begin{threeparttable}
 \renewcommand\arraystretch{1.6} 
\setlength{\tabcolsep}{0.5 em}{ 
    \begin{tabular}{c | c | c c | c c} 
    \hline\hline
    \multirow{3}{*}{ \fontsize{10}{12}\selectfont \text{Coupling constant} }& \multirow{3}{*}{\fontsize{10}{12}\selectfont \text{Structure of }$X_0(2900)$ }  &  \multicolumn{4}{c}{\fontsize{10}{12}\selectfont \text{Model}}\\ \cline{3-6}
    &   &  \multicolumn{2}{c|}{\fontsize{8}{12}\selectfont $a/(b+Q^2)$}&  \multicolumn{2}{c}{\fontsize{8}{12}\selectfont $a \;\mathrm{Exp}(-b\; Q^2)$}\\ \cline{3-6}
    &   & $a[\mathrm{GeV}^2]$ & $b[\mathrm{GeV}^2]$ & $a[\mathrm{GeV}^2]$ & $b[\mathrm{GeV}^{-2}]$ \\
    \hline
\multirow{3}{*}{$g_{\bar{D}KX_0}$}   
    & $\bar{D}^\ast K^\ast\;\mathrm{Molecule}$ & $25.56\pm 1.63$ & $57.43\pm0.00$ & - & - \\
    & $\mathrm{S-S}\;\mathrm{Tetraquark}$ & $18.20\pm 1.97$ & $57.43\pm0.00$ & - & - \\
    & $\mathrm{A-A}\;\mathrm{Tetraquark}$ & $22.20\pm 3.94$ & $57.43\pm0.00$ & - & - \\
    \hline
\multirow{3}{*}{$g_{\bar{D}^\ast K^\ast X_0}$}  
    & $\bar{D}^\ast K^\ast\;\mathrm{Molecule}$ & - & - & $0.12_{-0.008}^{+0.008}$ & $0.22\pm0.00$ \\
    & $\mathrm{S-S}\;\mathrm{Tetraquark}$ & - & - & $0.04_{-0.007}^{+0.015}$ & $0.22\pm0.00$ \\
    & $\mathrm{A-A}\;\mathrm{Tetraquark}$ & - & - & $0.04_{-0.014}^{+0.031}$ & $0.22\pm0.00$ \\
\hline
\multirow{3}{*}{$g_{\bar{D}_1 K_1 X_0}$}  
    & $\bar{D}^\ast K^\ast\;\mathrm{Molecule}$ & - & - & $0.38\pm0.01$ & $0.03^{+0.05}_{-0.02}$ \\
    & $\mathrm{S-S}\;\mathrm{Tetraquark}$ & - & - & $0.10\pm0.03$ & $0.03\pm0.01$ \\
    & $\mathrm{A-A}\;\mathrm{Tetraquark}$ & - & - & $0.20\pm0.02$ & $0.03\pm0.01$ \\
\hline
\multirow{3}{*}{$g_{\bar{D}_0^\ast K_0^\ast X_0}$}  
    & $\bar{D}^\ast K^\ast\;\mathrm{Molecule}$ & $13.59_{-6.21}^{+5.07}$ & $46.23\pm0.00$ & - & -  \\
    & $\mathrm{S-S}\;\mathrm{Tetraquark}$  & $2.11_{-0.63}^{+0.44}$ & $51.19\pm0.07$ & - & -\\
    & $\mathrm{A-A}\;\mathrm{Tetraquark}$ &  $3.86_{-0.90}^{+0.92}$ & $46.23\pm0.01$ & - & -\\    
     \hline\hline
    \end{tabular}
    }
    \label{Tab:FitcoefDKX}
  \end{threeparttable}
\end{table*}
\subsection{Coupling constant $g_{D_{s(0,1)}^{(\ast)} D K^{(\ast)}_{(0,1)}}$}
\par In this subsection, we consider the other strong coupling vertex in Fig.~\ref{Fig:BDXhadronlevel}. The interpolating currents for the relevent mesons are
\begin{equation}
\begin{split}
J_{D_{s}}=\mathrm{i}\bar{s}\gamma_5 c, &\quad J_{D_{s0}^\ast}=\bar{s}c,\\
J_{D_{s}^\ast}=\bar{s}\gamma_\mu c, &\quad J_{D_{s1}}=\bar{s}\gamma_\mu\gamma_5 c,\\
J_{D^+}&=\mathrm{i}\bar{d}\gamma_5 c.\\
\end{split}
\end{equation}
They can couple to the corresponding mesons via the following relations
\begin{equation}
\begin{split}
\langle0|J_{D_{s}}|D_{s}\rangle=f_{D_s}\frac{m_{D_s}^2}{m_c}\equiv\lambda_{D_{s}}, &\quad \langle0|J_{D_{s0}^\ast}|D_{s0}^\ast\rangle=f_{D_{s0}}\frac{m_{D_{s0}}^2}{m_c}\equiv\lambda_{D_{s0}^\ast},\\
\langle0|J_{D_{s}^\ast\mu}|D_{s}^\ast(p,\epsilon)\rangle=m_{D_{s}^\ast}f_{D_{s}^\ast}\epsilon_\mu\equiv\lambda_{D_{s}^\ast}\epsilon_\mu, &\quad \langle0|J_{D_{s1}\mu}|D_{s1}(p^{'},\epsilon)\rangle=m_{D_{s1}}f_{D_{s1}}\epsilon_\mu\equiv\lambda_{D_{s1}}\epsilon_\mu,\\
\langle0|J_{D^+}|D^+\rangle&=f_D\frac{m_D^2}{m_c}\equiv\lambda_{D}.
\end{split}
\end{equation}
The coupling constant $g_{D_{s(0,1)}^{(\ast)} D^{(\ast)}_{(0,1)}K^{(\ast)}_{(0,1)}}$ in Eqs.\eqref{Eq:AmpB-Ds1D-DX}-\eqref{Eq:AmpB-Ds1D0-DX} are defined via the effective Lagrangian~\cite{Casalbuoni:1992gi,Casalbuoni:1992dx,Casalbuoni:1996pg}
\begin{eqnarray}
\nonumber\mathcal{L}_{D_{s}^\ast DK}&=& -\mathrm{i}g_{D_s^\ast DK}\left(D\partial^\mu K D^{\ast\dagger}_{s\mu}-D_{s\mu}^\ast\partial^\mu K D^{\dagger}\right),\\
\nonumber\mathcal{L}_{D_{s1}DK^\ast}&=& g_{D_{s1}DK^\ast} D_{s1}^{\mu} K^{\ast}_\mu D^{\dagger} ,\\
\nonumber\mathcal{L}_{D_{s}DK^\ast}&=&\mathrm{i}g_{D_{s}DK^\ast}D\overset{\leftrightarrow}{\partial}_\mu D_s^\dagger K^{\ast\mu},\\
\nonumber\mathcal{L}_{D_{s0}DK}&=&g_{D_{s0}DK}\partial^\mu D\partial_\mu K,\\
\mathcal{L}_{D_{s}^\ast DK_1}&=&g_{D_{s}^\ast DK_1} DK_1^\mu D_{s\mu}^{\ast\dagger},\\
\nonumber\mathcal{L}_{D_{s0}DK_1}&=&\mathrm{i}g_{D_{s0}DK_1}D_{s0}^\ast \partial^\mu D K_{1\mu},\\
\nonumber\mathcal{L}_{D_{s}DK^\ast_0}&=&g_{D_{s}DK^\ast_0}K_0^\ast\partial_\mu D_s\partial^\mu D,\\
\nonumber\mathcal{L}_{D_{s1}DK^\ast_0}&=&\mathrm{i}g_{D_{s1}DK^\ast_0}DK_0^{\ast\mu}D^{\dagger}_{s1\mu},
\end{eqnarray}
thus the transition matrix element can be obtained as
\begin{eqnarray}
\nonumber\langle D(p^{'})K(q)|D_s^\ast(p)\rangle&=&\mathrm{i}g_{D_s^\ast DK} \epsilon_{D_s^\ast}\cdot q,\\
\nonumber\langle D(p^{'})K^\ast(q)|D_{s1}(p)\rangle&=&g_{D_{s1}DK^\ast} \epsilon_{K^\ast}^\ast\cdot \epsilon_{D_{s1}},\\
\nonumber\langle D(p^{'})K^\ast(q)|D_s(p)\rangle&=&\mathrm{i}g_{D_sDK^\ast} \epsilon_{K^\ast}\cdot (p+p^{'}),\\
\langle D(p^{'})K(q)|D_{s0}(p)\rangle&=&g_{D_{s0}DK} \;p^{'}\cdot q,\\
\nonumber\langle D(p^{'})K_1(q)|D_s^\ast(p)\rangle&=&g_{D_s^\ast DK_1} \epsilon^\ast_{D_s^\ast}\cdot \epsilon_{K_1},\\
\nonumber\langle D(p^{'})K_1(q)|D_{s0}^\ast(p)\rangle&=&\mathrm{i}g_{D_{s0}DK_1} \epsilon_{K_1}\cdot p^{'},\\
\nonumber\langle D(p^{'})K_0^\ast(q)|D_s(p)\rangle&=&g_{D_s DK_0^\ast}\; p\cdot p^{'},\\
\nonumber\langle D(p^{'})K^\ast_0(q)|D_{s1}(p)\rangle&=&\mathrm{i}g_{D_{s1}DK_0^\ast} \epsilon _{D_{s1}}^\ast\cdot q.
\end{eqnarray}
With the above coupling relations and transition matrix element, we can obtain the three-point correlator process on the phenomenological side. For $D_{s}^\ast DK$ process,
\begin{equation}
\begin{split}
\Pi_\mu(p,p^{'},q)=&\int\mathrm{d}^4x\mathrm{d}^4y\;\mathrm{e}^{\mathrm{i}p^{'}\cdot x}\mathrm{e}^{\mathrm{i}q\cdot y}\langle0|T\{J_{D}(x)J_{K^0}(y)J^\dagger_{D_s^\ast\mu}(0)\}|0\rangle\\
=&\frac{\lambda_{D_s^\ast}\lambda_{D}\lambda_{K}g_{D_s^\ast DK}}{(p^{'2}-m_{D}^2)(p^2-m_{D_s^\ast}^2)(q^2-m_K^2)}\left(p_\mu^{'}-\frac{m_{D_s^\ast}^2+m_D^2-q^2}{2m_{D_s^\ast}^2}p_\mu\right)+\cdots.
\end{split}
\end{equation}
\begin{figure}[bp]
\centering
\includegraphics[width=\textwidth]{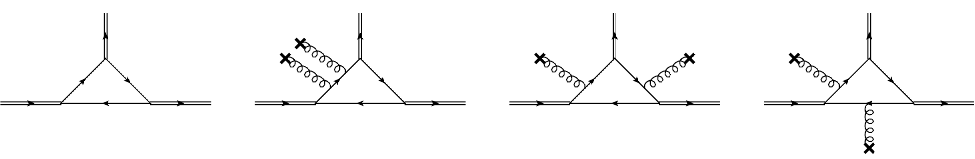}
\caption{The Feynman diagrams for OPE series of the correlation function \eqref{Eq:rhoDsDK}. The double solid line denotes the in(out)-coming hadron, while the single solid line denotes the quark propagator. Diagrams related by symmetry are not shown.}
\label{Fig:FeynDsDK}
\end{figure}
On the OPE side, we can evaluate the correlation function with the standard QCD sum rules approach. The related Feynman diagrams of the OPE series are shown in Fig.~\ref{Fig:FeynDsDK}. To establish a sum rule for the coupling constant $g_{D_s^\ast DK}$, we will pick out the $p_\mu$ structure in the OPE series and then match both sides of the sum rule. After performing the double Borel transform $P^2=-p^2\rightarrow M_B^2, P^{'2}=-p^{'2}\rightarrow M_B^{'2}$ on both sides, we obtain
\begin{equation}\label{Eq:gDsDK}
g_{D_s^\ast DK}(M_B^2,M_B^{'2},Q^2)=-\frac{1}{4\pi^2}\frac{2m_{D_s^\ast}^2}{\lambda_{D_s^\ast}\lambda_{D}\lambda_{K}\mathrm{e}^{-m_{D_s^\ast}^2/M_B^2-m_D^2/M_B^{'2}}}\frac{Q^2+m_K^2}{m_{D_s^\ast}^2+m_D^2+Q^2}\int_{m_c^2}^{s_0}\mathrm{d}s\int_{m_c^2}^{u_0}\mathrm{d}u\;\rho(s,u,t)\;\mathrm{e}^{-s/M_B^2-u/M_B^{'2}},
\end{equation}
where 
\begin{equation}\label{Eq:rhoDsDK}
\rho(s,u,t)=-\frac{48\pi^4tu(2m_c^2-s+t-u)}{\lambda^{3/2}(s,u,t)}+\frac{2\pi^4(s-t-3u)}{3\lambda^{3/2}(s,u,t)}\langle GG\rangle,
\end{equation}
and the Mandalstam invariances satisfy the relation:
\begin{equation}
 (s-m_c^2)(u-m_c^2)\geq Q^2m_c^2.
\end{equation}
\par
\begin{figure}[htbp]
\centering
\includegraphics[width=8cm]{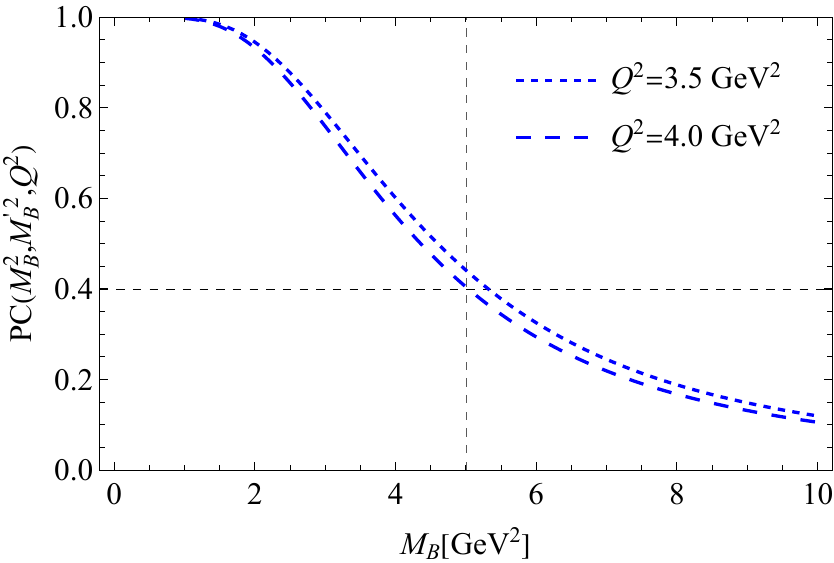}\quad
\includegraphics[width=8cm]{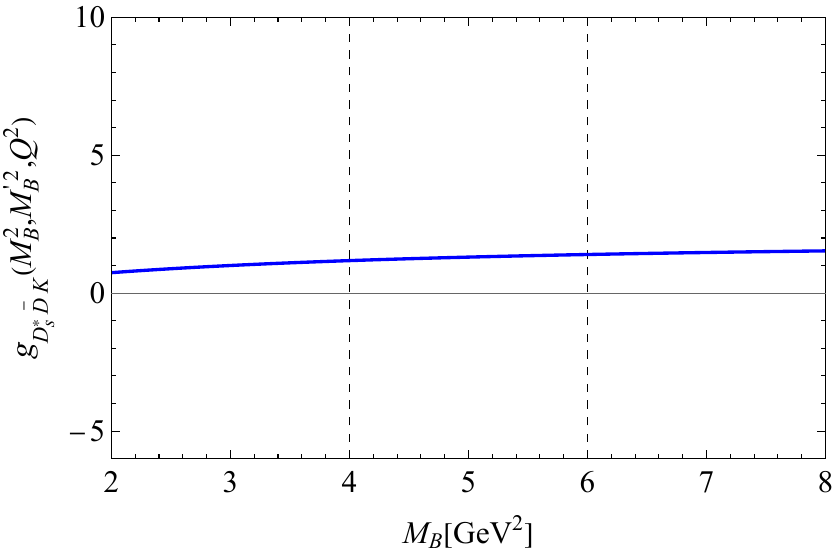}\\
\caption{The dependence of the pole contribution (left panel) and coupling $g_{D_s^\ast\bar{D}K}$ (right panel) on the Borel mass $M_B^2$. On the left panel, the transfer momentum is set to be $Q^2=3.5\;\mathrm{GeV}^2$ (blue dotted curve) and $Q^2=4.0\;\mathrm{GeV}^2$ (blue dashed curve), where the $Q^2$ region is chosen around $m_D^2$ and sufficient far away from $m_K^2$ to ensure the validity of sum rule. On the right panel, the coupling $g_{D_s^\ast\bar{D}K}$ with $Q^2=3.5\;\mathrm{GeV}^2$ behaves very stable around $M_B^2=5\;\mathrm{GeV}^2$,  with uncertainty estimated by the Borel mass uncertainty $\Delta M_B^2=\pm1\;\mathrm{GeV}^2$}.
\label{Fig:DsDKPC&MB}
\end{figure}

\begin{figure}[htbp]
\centering
\includegraphics[width=\textwidth]{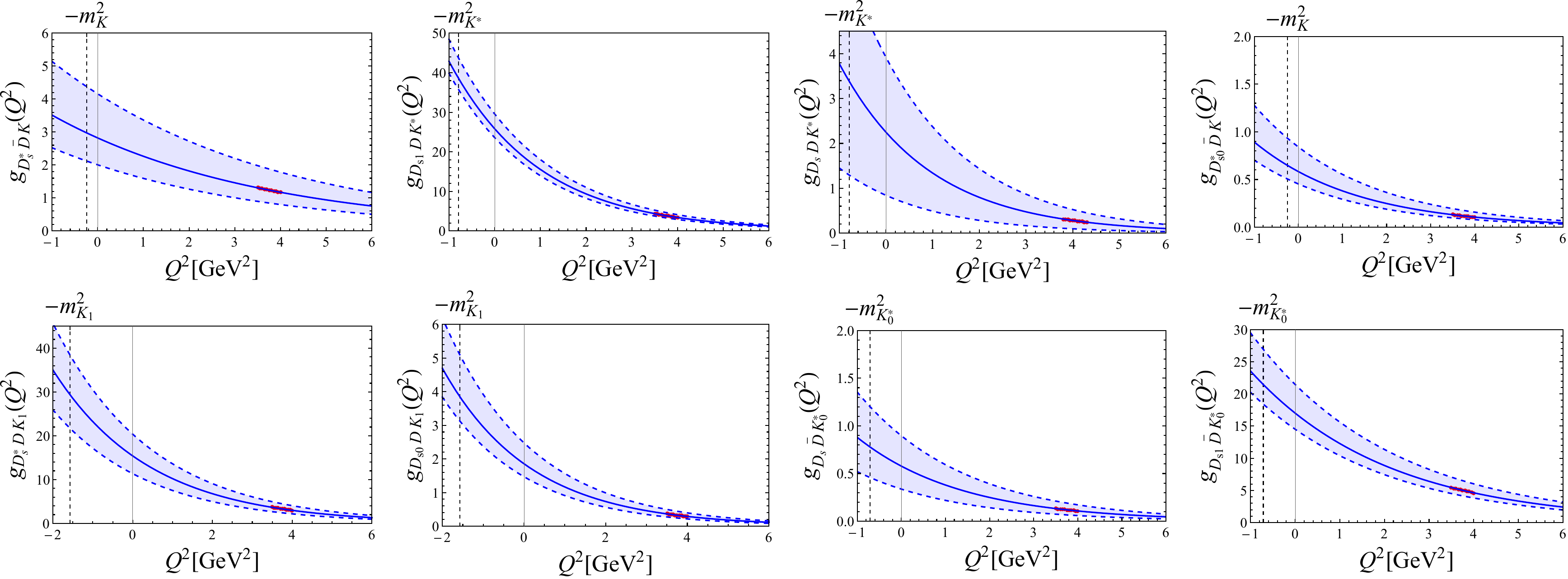}\\
\caption{The dependence of the coupling constant $g_{D_{s(0,1)}^{\ast}DK^{(\ast)}_{(0,1)}}$ on the transfer momentum $Q^2$, with Borel mass $M_B^2$ and continuum threshold $s_0$ listed in Tab.~\ref{Tab:DsDK}. For each figure, the red dots denote the value from the corresponding sum rules while the blue solid line is the exponential fitting curve. The two dashed blue lines denote the upper and lower boundary of the uncertainty from the Borel masses, various condensates, quark masses and hadronic parameters.}
\label{Fig:gQ2}
\end{figure}

 To perform the numerical analysis, we use the following relation between the Borel masses which is proposed in Ref.~\cite{Navarra:2000ji}:
\begin{equation}
M_B^{'2}=\frac{m_{\mathrm{out}}^2}{m_{\mathrm{in}}^2}M_B^2.
\end{equation}
In this case, we shall use the relation as follow:
\begin{equation}
M_B^{'2}=\frac{m_D^2}{m_{D_s^\ast}^2}M_B^2,
\end{equation}
and the continuum threshold parameters is taken from Ref.~\cite{Gelhausen:2013wia}. To determine the proper Borel masses in our sum rule analysis, we define the pole contribution(PC) as
\begin{equation}
\mathrm{PC}(Q^2)=\frac{\int_{m_c^2}^{s_0}\mathrm{d}s^{'}\int_{m_c^2}^{u_0}\mathrm{d}u^{'}\rho(s^{'},u^{'},-Q^2)\;\mathrm{e}^{-s^{'}/M_B^2-u^{'}/M_B^{'2}}}{\int_{m_c^2}^{\infty}\mathrm{d}s^{'}\int_{m_c^2}^{\infty}\mathrm{d}u^{'}\rho(s^{'},u^{'},-Q^2)\;\mathrm{e}^{-s^{'}/M_B^2-u^{'}/M_B^{'2}}}
\end{equation}
and we ensure the PC at least at 40\%. We show the PC for $g_{D_s^\ast DK}$ in the left panel of Fig.~\ref{Fig:DsDKPC&MB} and find that the proper $M_B^2$ should be $5.0\;\mathrm{GeV}^2$. We show the dependence of $g_{D_s^\ast DK}$ on Borel mass in the right panel of Fig.~\ref{Fig:DsDKPC&MB} and we can find a relatively steady platform around the chosen Borel mass, where we can obtain a stable result at $Q^2=3.5\;\mathrm{GeV}^2\sim m_D^2$.
We fit our result from $Q^2$ range 3.5 to 4.0 GeV$^2$ with the exponential model:
\begin{equation}
g_{D_s^\ast DK}(Q^2)=(2.82^{+1.34}_{-0.82}\;\mathrm{GeV}^{-2})\;\mathrm{e}^{-(-0.22\pm0.01\mathrm{GeV}^{-2})Q^2},
\end{equation}
which is shown in Fig.~\ref{Fig:gQ2}. The numerical analysis for coupling constants of other vertices is similar, we only list out their corresponding sum rules and spectrum functions in Appendix B and list out the results in Tab.~\ref{Tab:DsDK}. The exponential fitting curves are also shown in Fig.~\ref{Fig:gQ2}.


\begin{table}[htbp]
\caption{Fitting coefficients with exponential model of coupling constants. The continuum thresholds $s_0$ are taken from Ref.~\cite{Gelhausen:2013wia}. The latter two columns shows the coupling constants with on-shell condition $Q^2=-m_{K_{(0,1)}^{(\ast)}}^2$ in this work and other works.}\label{Tab:DsDK}\renewcommand\arraystretch{1.6} 
\setlength{\tabcolsep}{0.5 em}{ 
\centering
\begin{tabular}{c |c c c c| c c}
  \hline
 \hline
                  & $M_B^2[\mathrm{GeV}^{2}]$ & $s_0[\mathrm{GeV}^{2}]$ & $a[\mathrm{GeV}^{2}]$ & $b[\mathrm{GeV}^{-2}]$ & $g$ & Other Work \\
 \hline
$g_{D_s^\ast DK}$ & 5.0 & 7.40  & $2.82^{+1.34}_{-0.82}$ &  $0.22\pm0.00$   &  $2.98_{-1.61}^{+1.90}$   &  $2.84\pm0.31$~\cite{Bracco:2006xf} \\
$g_{D_{s0}^\ast DK}$ & 8.6 & 6.84  & $0.58^{+0.26}_{-0.13}$ &  $0.43\pm0.01$   &  $0.65_{-0.14}^{+0.29}$   &  $0.52\pm0.17$~\cite{Colangelo:1995ph} \\
$g_{D_{s1} DK^\ast}$ & 4.8 & 6.25  & $25.81^{+6.60}_{-3.45}$ &  $0.51\pm0.01$   &  $38.59_{-3.66}^{+5.70}$   &  $23.7\sim24.9$~\cite{Lin:2024hys} \\
$g_{D_s DK^\ast}$ & 5.0 & 6.30  & $2.25^{+1.67}_{-1.41}$ &  $0.52^{+0.03}_{-0.02}$   &  $3.39_{-2.19}^{+2.56}$   & $2.85\pm0.46$~\cite{Janbazi:2017mpb}  \\
$g_{D_s^\ast DK_1}$ & 5.0 & 7.40  & $15.41^{+4.96}_{-4.06}$ &  $0.41\pm0.00$   &  $29.36_{-7.84}^{+9.74}$   &  $11.92\pm4.64$~\cite{Momeni:2020scc} \\
$g_{D_{s0}^\ast DK_1}$ & 5.2 & 6.84  & $1.85^{+0.62}_{-0.39}$ &  $0.46\pm0.01$   &  $3.84_{-0.89}^{+1.32}$   & -  \\
$g_{D_s DK_0^\ast}$ & 6.4 & 6.30  & $0.58^{+0.32}_{-0.24}$ &  $0.42\pm0.01$   &  $0.78_{-0.33}^{+0.43}$   & $0.74\pm0.05$~\cite{Sundu:2011vz}  \\
$g_{D_{s1} DK_0^\ast}$ & 5.0 & 6.25  & $17.01^{+4.45}_{-2.50}$ &  $0.33\pm0.00$   &  $21.45_{-3.28}^{+5.66}$   &  - \\
 \hline
  \hline
\end{tabular}}
\end{table}
\section{Decay and production of $X(2900)$}\label{Sec.Production&Decay}
\par We obtain all the coupling constants in Eq.~\eqref{Eq:AmpB-DsD0-DX} and Eq.~\eqref{Eq:AmpB-Ds1D-DX} in the above sections, in this section we shall obtain the branching fraction of the $B\rightarrow DX_0(2900)$ production process as well as the decay width of the $X_0(2900)$. The decay width of the two-body strong decay process $X_0(2900)\rightarrow\bar{D}K^0$ can be derived from Eq.\eqref{Eq:MatrixelementofDKX} as
\begin{equation}
\Gamma(X_0(2900)\rightarrow\bar{D}^0K^0)=\frac{\lambda^{1/2}(m_{X_0}^2,m_D^2,m_K^2)}{64\pi m_{X_0}}g^{2}_{\bar{D}KX_0}(-m_K^2)\left(m_{X_0}^2-m_D^2-m_K^2\right)^2.
\end{equation}
Then we can calculate the partical decay width of the process $X_0(2900)\rightarrow\bar{D}K^0$ as
\begin{equation}
\Gamma(X_0(2900)\rightarrow\bar{D}^0K^0)=32.03\pm9.12\;\mathrm{MeV},
\end{equation}
where the error mainly comes from the uncertainties of the coupling constant $g_{\bar{D}KX_0}$. It should be noted that in our previous calculation in Sec.\ref{Sec.QCDSR}, we adopt the light quark mass as $m_q=0$, which would lead to the isospin symmetry $g_{\bar{D}^0K^0X_0}=g_{D^-K^+X_0}$. Ignoring the small differences due to phase space, the total $\bar{D}K$ decay width should be
\begin{equation}
\begin{split}
\Gamma(X_0(2900)\rightarrow\bar{D}K)&=\Gamma(X_0(2900)\rightarrow\bar{D}^0K^0)+\Gamma(X_0(2900)\rightarrow D^-K^+)\\
&=2\Gamma(X_0(2900)\rightarrow\bar{D}^0K^0)\\
&=64.07\pm18.23\;\mathrm{MeV}.
\end{split}
\end{equation}
Due to the estimations that $X_0(2900)$ dominantly decay into $\bar{D}K$ channel~\cite{Huang:2020ptc,Xiao:2020ltm}, our result is well consistent with the experimental result Eq.\eqref{Eq:ExpofX2900}. The decay widths for other structures of $X_0(2900)$ are listed in Tab.~\ref{Tab:BrFr&Decaywidth}.
\begin{table}[htbp]
\caption{The values of the parameters $F(0),a$ and $b$ in the form factors Eq.\eqref{Eq:FormFactorPtrized} for $B\rightarrow D^{(\ast)}_{(0,1)}$ taken from Ref.~\cite{Cheng:2003sm}.}\label{Tab:FFBDs}\renewcommand\arraystretch{1.6} 
\setlength{\tabcolsep}{0.5 em}{ 
\centering
\begin{tabular}{c c c c | c c c c}
  \hline
 \hline
Form Factor & $F(0)$ & $a$ & $b$ & Form Factor & $F(0)$ & $a$ & $b$ \\
\hline
$F_1^{B\rightarrow D}$ & 0.67 & 1.22 & 0.36 & $F_1^{B\rightarrow D_0^\ast}$ & 0.24 & 1.03 & 0.27 \\
$F_0^{B\rightarrow D}$ & 0.67 & 0.63 & 0.01 & $F_0^{B\rightarrow D_0^\ast}$ & 0.24 & -0.49 & 0.35 \\
$A_0^{B\rightarrow D^\ast}$ & 0.68 & 1.21 & 0.36 & $A_0^{B\rightarrow D_1}$ & 0.08 & 1.28 & -0.29 \\
$A_1^{B\rightarrow D^\ast}$ & 0.65 & 0.60 & 0.00 & $A_1^{B\rightarrow D_1}$ & -0.19 & -1.25 & 0.97 \\
$A_2^{B\rightarrow D^\ast}$ & 0.61 & 1.12 & 0.31 & $A_2^{B\rightarrow D_1}$ & -0.12 & 0.67 & 0.20 \\
$A_V^{B\rightarrow D^\ast}$ & 0.77 & 1.25 & 0.38 & $A_V^{B\rightarrow D_1}$ & -0.12 & 0.71 & 0.18 \\
 \hline
  \hline
\end{tabular}}
\end{table}
\begin{table}[htbp]
\caption{Amplitudes for diagrams in Fig.~\ref{Fig:BDXhadronlevel}, branch fraction and decay width for $X_0(2900)$ with molecule structure and tetraquark structure. The subscription indices a-h denote the corresponding diagrams in Fig.~\ref{Fig:BDXhadronlevel}. The latter three columns denote the molecule structure, scalar-scalar compact structure and axialvector-axialvector structure correspondingly. }\label{Tab:BrFr&Decaywidth}\renewcommand\arraystretch{1.6} 
\setlength{\tabcolsep}{0.5 em}{ 
\centering
\begin{tabular}{c| c c c}
  \hline
 \hline
 & mol & S-S  & A-A\\ 
 \hline
$\mathcal{A}_a[\times10^{-8}]$ & $-4.16_{-2.50}^{+1.71}$ & $-2.97^{+1.35}_{-1.91}$ & $-3.62_{-2.26}^{+1.57}$ \\
$\mathcal{A}_b[\times10^{-8}]$ & $+3.67^{+0.78}_{-0.34}$ & $+1.14^{+0.63}_{-0.31}$ & $+1.08^{+0.62}_{-0.30}$ \\
$\mathcal{A}_c[\times10^{-8}]$ & $+0.10^{+0.04}_{-0.05}$ & $+0.03_{-0.02}^{+0.03}$ & $+0.03^{+0.03}_{-0.02}$ \\
$\mathcal{A}_d[\times10^{-8}]$ & $-4.33^{+0.81}_{-2.15}$ & $-3.09_{-1.67}^{+0.72}$ & $-3.77_{-1.96}^{+0.80}$ \\
$\mathcal{A}_e[\times10^{-8}]$ & $-1.73_{-0.43}^{+0.31}$ & $-0.45^{+0.08}_{-0.24}$ & $-0.90_{-0.35}^{+0.16}$ \\
$\mathcal{A}_f[\times10^{-8}]$ & $+0.02^{+0.01}_{-0.00}$ & $+0.01^{+0.04}_{-0.00}$ & $+0.01_{-0.01}^{+0.05}$ \\
$\mathcal{A}_g[\times10^{-8}]$ & $-0.91_{-0.02}^{+0.05}$ & $-0.13^{+0.09}_{-0.08}$ & $-0.26_{-0.14}^{+0.17}$ \\
$\mathcal{A}_h[\times10^{-8}]$ & $-0.57_{-0.07}^{+0.07}$ & $-0.08_{-0.03}^{+0.03}$ & $-0.16^{+0.05}_{-0.04}$ \\
\hline
$\mathcal{B}r(B^+\rightarrow D^+X_0)[\times10^{-5}]$   & $2.47^{+2.07}_{-1.55}$ & $1.20^{+1.39}_{-0.80}$ & $2.26^{+2.06}_{-1.33}$ \\
$\Gamma(X_0\rightarrow DK)[\mathrm{MeV}]$ & $64.07\pm18.23$ & $32.49\pm21.11$ & $48.31\pm21.78$\\
 \hline
  \hline
\end{tabular}}
\end{table}

\par With the coupling constants obtained in Sec.~\ref{Sec.QCDSR} and the weak transition form factor is parametrized as~\cite{Cheng:2003sm}
\begin{equation}\label{Eq:FormFactorPtrized}
F(Q^2)=\frac{F(0)}{1-a(Q^2/m_B^2)+b(Q^2/m_B^2)^2},
\end{equation}
the corresponding parameters are listed in Tab.~\ref{Tab:FFBDs}, we can calculate the amplitude \eqref{Eq:AmpB-DsD0-DX}-\eqref{Eq:AmpB-Ds1D0-DX} as well as the decay width of the production process $B^+\rightarrow D^+X_0(2900)$ through
\begin{equation}
\Gamma(B^+\rightarrow D^+X_0)=\frac{\lambda^{1/2}(m_B^2,m_D^2,m_{X_0}^2)}{16\pi m_B^3}\left|\sum_{i=a}^h\mathcal{A}_i\right|^2,
\end{equation}
and its corresponding branching fraction
\begin{equation}
\mathcal{B}r(B^+\rightarrow D^+X_0)=\frac{\Gamma(B^+\rightarrow D^+X_0)}{\Gamma_{\mathrm{tot}}}.
\end{equation}
We list out the results in Tab.~\ref{Tab:BrFr&Decaywidth}.
\section{Discussion and conclusion}\label{Sec.Dis&Con}
\par In this work, we mainly consider the branching fraction of the $X_0(2900)$ production process shown in Fig.~\ref{Fig:BDXhadronlevel} with different interpretation of $X_0(2900)$ state. By deriving the decay amplitudes of Fig.~\ref{Fig:BDXhadronlevel}, we investigate the related strong decay vertices with QCD sum rule approach. The corresponding results are listed in Tab.~\ref{Tab:BrFr&Decaywidth}, from which one can see that the branching fraction and decay width of $X_0(2900)$ in the molecular picture are well consistent with the experimental results~\cite{LHCb:2020bls,LHCb:2020pxc}, indicating that the $X_0(2900)$ state could be more likely a molecular state candidate. Such conclusion meets agreement with Ref.~\cite{Hu:2020mxp,Kong:2021ohg,Ke:2022ocs,Liu:2020nil,Wang:2021lwy,Chen:2020aos,Agaev:2020nrc,Chen:2021erj,Mutuk:2020igv,Huang:2020ptc,Xiao:2020ltm}, where the decay width is calculated as $59.37^{+24.94}_{-17.96}\;\mathrm{MeV}$ with effective Lagrangian formalism~\cite{Huang:2020ptc} and $49.6\pm9.3\;\mathrm{MeV}$ with light-cone QCD sum rule~\cite{Agaev:2020nrc}. But we still cannot exclude the interpretation for axialvector-axialvector compact structure within the uncertainty. From Tab.~\ref{Tab:BrFr&Decaywidth}, we can see that the major contributions come from the $B^+\rightarrow \bar{D}^0D_s^{\ast+}\rightarrow D^+X_0(2900)$, $B^+\rightarrow \bar{D}^{\ast0}D_{s1}^{+}\rightarrow D^+X_0(2900)$ and $B^+\rightarrow \bar{D}^0D_{s0}^{+}\rightarrow D^+X_0(2900)$ processes, while the contribution from $B^+\rightarrow \bar{D}^{\ast0}D_{s1}^{+}\rightarrow D^+X_0(2900)$ process in the molecular interpretation is much larger than that in the compact interpretations. It is because the coupling from $\bar{D}^{\ast}K^\ast X_0$ vertex in the molecular interpretation is much larger than that in the compact one. It is interesting to note that our branching fraction of the process $B^+\rightarrow D^+X\rightarrow D^+(D^-K^+)$, $B^+\rightarrow D^+X\rightarrow D^+(\bar{D}^0K^0)$, $B^0\rightarrow D^0X\rightarrow D^0(\bar{D}^0K^0)$ and $B^0\rightarrow D^0X\rightarrow D^0(D^-K^+)$ should be the same due to $\mathcal{B}r(B\rightarrow DX\rightarrow D\bar{D}K)=\mathcal{B}r(B\rightarrow DX)\mathcal{B}r(X\rightarrow \bar{D}K)$.
In Ref.~\cite{Burns:2020xne}, the result shows that the branching fraction of process $B^+\rightarrow D^+X\rightarrow D^+(\bar{D}^0K^0)$ and $B^0\rightarrow D^0X\rightarrow D^0(D^-K^+)$ would be highly suppressed with the triangle model. Future experimental studies can discriminate among the molecule, compact tetraquark or triangle singularity interpretations for the $X_0(2900)$ state.
\section*{ACKNOWLEDGMENTS}
This work is partly supported by the National Natural Science Foundation of China with Grant Nos.~12375073, 12035007, and 12175318, Guangdong Provincial funding with Grant Nos.~2019QN01X172, Guangdong Major Project of Basic and Applied Basic Research No.~2020B0301030008, the Natural Science Foundation of Guangdong Province of China under Grant No. 2022A1515011922, 
 the NSFC and the Deutsche Forschungsgemeinschaft (DFG, German
Research Foundation) through the funds provided to the Sino-German Collaborative
Research Center TRR110 ``Symmetries and the Emergence of Structure in QCD"
(NSFC Grant No. 12070131001, DFG Project-ID 196253076-TRR 110).
\appendix
\section{Spectrum function for three-point correlation of $D_{(0,1)}^{(\ast)}K_{(0,1)}^{(\ast)}X_0$ vertices}
\par The spectrum function $\rho(s)$ and $R(M_B^2)$ in Eq.\eqref{Eq:gDKX} with $\bar{D}^\ast K^\ast$ molecule structure is shown as follow
\begin{eqnarray}
\rho(s)&=&\int_{x_{\mathrm{min}}}^{1}\mathrm{d}x\frac{m_c(1-x)}{8\pi^2}(\qq+\qss)+\int_0^1\mathrm{d}x\left(\frac{m_cm_s(x-1)}{256\pi^4}\GG-\frac{m_cx(x-1)}{32\pi^2}(\qGq+\sGs)\right),\\
\nonumber R(M_B^2)&=&-\frac{1}{96\pi^2M_B^8}\left(2m_c^2m_s\qq\GG+\pi^2M_B^6\qGq(\qq+\qss)+4M_B^4\left(3m_s\qGq+16\pi^2\qq(\qq+\qss)\right)\right.\\
& &\left.+4M_B^2m_c^2\left(3m_s\qGq+16\pi^2\qq(\qq+\qss)\right)\right),
\end{eqnarray}
where $x_{\mathrm{min}}=1-m_c^2/s$ and $\Delta(s,x)=(1-x)(m_c^2-sx)$, and in Eq.\eqref{Eq:gDsKsX}
\begin{eqnarray}
\nonumber\rho(s)&=&\int_{x_{\mathrm{min}}}^{1}\mathrm{d}x\left(-\frac{2m_s\Delta(s,x)+m_ss(x-1)x}{4\pi^2}\qq+\frac{m_s\Delta(s,x)+m_ss(x-1)x}{8\pi^2}\qss\right.\\
\nonumber& &\left.-\frac{3\Delta(s,x)+s(x-1)x}{384\pi^4}\GG-\frac{m_sx}{16\pi^2}(2\qGq+\sGs)\right)\\
& &+\int_0^1\mathrm{d}x\left(\frac{m_ss(1-x)x^2}{48\pi^2}(3\qGq+\sGs)+\frac{m_s(x-1)x}{192\pi^2}\GG(\qss-3\qq)\right),\\
R(M_B^2)&=&-\frac{1}{96\pi^2M_B^6}\left(4m_c^3\qq\GG+\pi^2m_cm_s(4\qq-\qss)(M_B^4\qGq+64m_c^2\qq)\right),
\end{eqnarray}
and in Eq.\eqref{Eq:gD1K1X}
\begin{eqnarray}
\nonumber\rho(s)&=&\int_{x_{\mathrm{min}}}^{1}\mathrm{d}x\left(-\frac{2m_s\Delta(s,x)+m_ss(x-1)x}{4\pi^2}\qq+\frac{m_s\Delta(s,x)+m_ss(x-1)x}{8\pi^2}\qss\right.\\
\nonumber& &\left.-\frac{3\Delta(s,x)+s(x-1)x}{384\pi^4}\GG+\frac{m_sx}{16\pi^2}(2\qGq-\sGs)\right)\\
& &+\int_0^1\mathrm{d}x\left(\frac{m_ss(1-x)x^2}{48\pi^2}(3\qGq-\sGs)+\frac{m_s(x-1)x}{192\pi^2}\GG(\qss+4\qq)\right),\\
R(M_B^2)&=&\frac{1}{96\pi^2M_B^6}\left(4m_c^3\qq\GG-\pi^2m_cm_s(4\qq+\qss)(M_B^4\qGq+64m_c^2\qq)\right),
\end{eqnarray}
and in Eq.\eqref{Eq:gD0K0X}
\begin{eqnarray}
\rho(s)&=&\int_{x_{\mathrm{min}}}^{1}\mathrm{d}x\frac{m_c(1-x)}{8\pi^2}(\qss-\qq)-\int_0^1\mathrm{d}x\left(\frac{m_c(1-x)x}{32\pi^2}(\qGq+\sGs)+\frac{m_cm_s(1-x)}{256\pi^4}\GG\right),\\
\nonumber R(M_B^2)&=&\frac{1}{96\pi^2M_B^8}\Bigg(2m_c^2m_s\qq\GG+\pi^2M_B^6\qGq(\qss-\qq)+4M_B^4\Big(3m_s\qGq+16\pi^2\qq(\qss-\qq)\\
& &+4M_B^2m_c^2(3m_s\qGq+16\pi^2\qq(\qss-\qq))\Big)\Bigg).
\end{eqnarray}
The spectum function $\rho(s)$ and $R(M_B^2)$ in Eq.\eqref{Eq:gDKX} with scalar-diquark-scalar-antidiquark structure is shown as follow
\begin{eqnarray}
\rho(s)&=&\int_{x_{\mathrm{min}}}^{1}\mathrm{d}x\frac{m_c(1-x)}{8\pi^2}(\qq+\qss),\\
R(M_B^2)&=&-\frac{1}{96M_B^6}(\qq+\qss)\left(M_B^2\qGq+64\qq(M_B^2+m_c^2)\right),
\end{eqnarray}
and in Eq.\eqref{Eq:gDsKsX}
\begin{eqnarray}
\nonumber\rho(s)&=&\int_{x_{\mathrm{min}}}^{1}\mathrm{d}x\left(\frac{2m_s(\Delta(s,x)+s(x-1)x)}{4\pi^2}\qq-\frac{m_s(\Delta(s,x)+s(x-1)x)}{8\pi^2}\qss+\frac{3\Delta(s,x)+s(x-1)x}{384\pi^4}\GG\right)\\
& &+\int_{0}^{1}\mathrm{d}x\left(\frac{m_s(x-1)x}{192\pi^2}\qss\GG-\frac{m_s(x-1)x}{48\pi^2}\qq\GG\right),\\
R(M_B^2)&=&\frac{1}{96\pi^2M_B^6}\left(4m_c^3\qq\GG+\pi^2m_cm_s(4\qq-\qss)\left(M_B^4\qGq+64m_c^2\qq\right)\right),
\end{eqnarray}
and in Eq.\eqref{Eq:gD1K1X}
\begin{eqnarray}
\nonumber\rho(s)&=&\int_{x_{\mathrm{min}}}^{1}\mathrm{d}x\left(\frac{2m_s\Delta(s,x)+m_ss(x-1)x}{4\pi^2}\qq+\frac{m_s\Delta(s,x)+m_ss(x-1)x}{8\pi^2}\qss-\frac{3\Delta(s,x)+s(x-1)x}{384\pi^4}\GG\right)\\
& &+\int_0^1\mathrm{d}x\frac{m_s(1-x)x^2}{192\pi^2}(4\qq+\qss)\GG,\\
R(M_B^2)&=&\frac{1}{96\pi^2M_B^6}\left(4m_c^3\qq\GG-\pi^2m_cm_s(4\qq+\qss)(M_B^4\qGq+64m_c^2\qq)\right),
\end{eqnarray}
and in Eq.\eqref{Eq:gD0K0X}
\begin{eqnarray}
\rho(s)&=&\int_{x_{\mathrm{min}}}^{1}\mathrm{d}x \frac{m_c(1-x)}{8\pi^2}(\qss-\qq)\\
R(M_B^2)&=&-\frac{1}{96 M_B^6}(\qq-\qss)\left(M_B^4\qGq+64\qq(M_B^2+m_c^2)\right).
\end{eqnarray}
The spectum function $\rho(s)$ and $R(M_B^2)$ in Eq.\eqref{Eq:gDKX} with axialvector-diquark-axialvector-antidiquark structure is shown as follow
\begin{eqnarray}
\rho(s)&=&\int_{x_{\mathrm{min}}}^{1}\mathrm{d}x\frac{m_c(x-1)}{4\pi^2}(\qq+\qss),\\
R(M_B^2)&=&\frac{1}{48M_B^6}(\qq+\qss)\left(64(M_B^2+m_c^2)\qq+M_B^4\qGq\right),
\end{eqnarray}
and in Eq.\eqref{Eq:gDsKsX}
\begin{eqnarray}
\nonumber\rho(s)&=&\int_{x_{\mathrm{min}}}^{1}\mathrm{d}x\left(\frac{2m_s\Delta(s,x)+m_ss(x-1)x}{2\pi^2}\qq-\frac{m_s\Delta(s,x)+m_ss(x-1)x}{4\pi^2}\qss+\frac{3\Delta(s,x)+s(x-1)x}{192\pi^4}\GG\right)\\
& &+\int^1_0\mathrm{d}x\frac{m_s(x-1)x}{24\pi^2}\GG(\qss-4\qq),\\
R(M_B^2)&=&\frac{1}{48\pi^2M_B^6}\left(4m_c^3\qq\GG+\pi^2m_cm_s(4\qq-\qss)\left(M_B^4\qGq+64m_c^2\qq\right)\right),
\end{eqnarray}
and in Eq.\eqref{Eq:gD1K1X}
\begin{eqnarray}
\nonumber\rho(s)&=&\int_{x_{\mathrm{min}}}^{1}\mathrm{d}x\left(-\frac{2m_s\Delta(s,x)+m_ss(x-1)x}{2\pi^2}\qq-\frac{m_s\Delta(s,x)+m_ss(x-1)x}{4\pi^2}\qss+\frac{3\Delta(s,x)+s(x-1)x}{192\pi^4}\GG\right)\\
& &-\int_0^1\mathrm{d}x\frac{m_s(1-x)x}{96\pi^2}(4\qq+\qss)\GG,\\
R(M_B^2)&=&\frac{1}{48\pi^2M_B^6}\left(-4m_c^3\qq\GG+\pi^2m_cm_s(4\qq+\qss)(M_B^4\qGq+64m_c^2\qq)\right),
\end{eqnarray}
and in Eq.\eqref{Eq:gD0K0X}
\begin{eqnarray}
\rho(s)&=&\int_{x_{\mathrm{min}}}^{1}\mathrm{d}x \frac{m_c(x-1)}{4\pi^2}(\qss-\qq)\\
R(M_B^2)&=&-\frac{1}{48 M_B^6}(\qq-\qss)\left(M_B^4\qGq+64\qq(M_B^2+m_c^2)\right),
\end{eqnarray}
\section{Sum rule for three-point correlation of $D_{s(0,1)}^{(\ast)}DK_{(0,1)}^{(\ast)}$ vertices}
\par For $D_{s1}DK^\ast$ process,
the sum rule can be obtained as
\begin{equation}\label{Eq:gDs1DKs}
g_{D_{s1}DK^\ast}(M_B^2,M_B^{'2},Q^2)=-\frac{1}{4\pi^2}\frac{4m_{D_{s1}}^2m_{K^\ast}^2}{\lambda_{D_{s1}}\lambda_{D}\lambda_{K^\ast}\mathrm{e}^{-m_{D_{s1}}^2/M_B^2-m_D^2/M_B^{'2}}}\frac{Q^2+m_K^2}{\lambda(m_D^2,m_{D_{s1}}^2,-Q^2)+12m_{D_{s1}}^2m_{K^\ast}^2}\int_{m_c^2}^{s_0}\mathrm{d}s^{'}\int_{m_c^2}^{u_0}\mathrm{d}u^{'}\rho(s^{'},u^{'},t)\;\mathrm{e}^{-s^{'}/M_B^2-u^{'}/M_B^{'2}},
\end{equation}
where 
\begin{equation}
\rho(s,u,t)=\frac{48\pi^4(2m_c^2m_s-m_c(s+u)-2m_s u)}{\lambda^{1/2}(s,u,t)}.
\end{equation}
\par For $D_{s0}^\ast DK$ process, the sum rule can be obtained as
\begin{equation}\label{Eq:gDs0DK}
g_{D_{s0}^\ast DK}(M_B^2,M_B^{'2},Q^2)=-\frac{1}{4\pi^2}\frac{1}{\lambda_{D_{s0}}\lambda_{D}\lambda_{K}\mathrm{e}^{-m_{D_{s0}}^2/M_B^2-m_D^2/M_B^{'2}}}\frac{2(Q^2+m_K^2)}{m_{D_{s0}}^2-m_D^2+Q2}\int_{m_c^2}^{s_0}\mathrm{d}s^{'}\int_{m_c^2}^{u_0}\mathrm{d}u^{'}\rho(s^{'},u^{'},t)\;\mathrm{e}^{-s^{'}/M_B^2-u^{'}/M_B^{'2}},
\end{equation}
where
\begin{equation}
\begin{split}
\rho(s,u,t)=&-\frac{48\pi^4}{\lambda^{5/2}(s,u,t)}\Bigg(2m_c^3\Big(m_c(m_c+m_s)-s+t-u\Big)(\lambda(s,u,t)+3t(s-t+u))+m_s u(s-t+u)\lambda(s,u,t)\\
&-m_c^2m_s\Big(s(s^2+s(t+u)-5(t-u)^2)+3(t-u)^2(t+u)\Big)+m_c\Big(s^2(s^2-(s-t-u)(3t+2u)-s(t-u)(t^2+3tu-2u^2)\\
&-u(t-u)^3\Big)+6m_sstu(s-t+u)\Bigg)-\frac{24\pi^4\GG m_c(s(t-2u)+t(u-t))}{\lambda^{5/2}(s,u,t)}.
\end{split}
\end{equation}
\par For $D_{s}DK^\ast$ process, the sum rule can be obtained as
\begin{equation}\label{Eq:gDsDKs}
g_{D_{s}DK^\ast}(M_B^2,M_B^{'2},Q^2)=-\frac{1}{4\pi^2}\frac{m_{K^\ast}^2}{\lambda_{D_{s}}\lambda_{D}\lambda_{K^\ast}\mathrm{e}^{-m_{D_{s}}^2/M_B^2-m_D^2/M_B^{'2}}}\frac{Q^2+m_{K^\ast}^2}{m_{D_{s}}^2-m_D^2-m_{K^\ast}^2}\int_{m_c^2}^{s_0}\mathrm{d}s^{'}\int_{m_c^2}^{u_0}\mathrm{d}u^{'}\rho(s^{'},u^{'},t)\;\mathrm{e}^{-s^{'}/M_B^2-u^{'}/M_B^{'2}},
\end{equation}
where
\begin{equation}
\rho(s,u,t)=-\frac{48\pi^4(s-t-u)(m_c^4-m_c^2(s-t+u)+su)}{\lambda^{3/2}(s,u,t)}+\frac{2\pi^4\GG (s-t-u)}{3\lambda^{3/2}.(s,u,t)}
\end{equation}
\par For $D_{s}^\ast DK_1$ process, the sum rule can be obtained as
\begin{equation}\label{Eq:gDssDK1}
g_{D_{s}^\ast DK_1}(M_B^2,M_B^{'2},Q^2)=-\frac{1}{4\pi^2}\frac{4m_{D_{s}^\ast}^2m_{K_1}^2}{\lambda_{D_{s}^\ast}\lambda_{D}\lambda_{K_1}\mathrm{e}^{-m_{D_{s}^\ast}^2/M_B^2-m_D^2/M_B^{'2}}}\frac{Q^2+m_K^2}{\lambda(m_D^2,m_{D_{s}^\ast}^2,-Q^2)+12m_{D_{s}^\ast}^2m_{K_1}^2}\int_{m_c^2}^{s_0}\mathrm{d}s^{'}\int_{m_c^2}^{u_0}\mathrm{d}u^{'}\rho(s^{'},u^{'},t)\;\mathrm{e}^{-s^{'}/M_B^2-u^{'}/M_B^{'2}},
\end{equation}
where
\begin{equation}
\rho(s,u,t)=-\frac{48\pi^4(2m_c^2m_s+m_c(s+u)-2m_su)}{\sqrt{\lambda(s,u,t)}}.
\end{equation}
\par For $D_{s0}^\ast DK_1$ process, the sum rule can be obtained as
\begin{equation}\label{Eq:gDs0DK1}
g_{D_{s0}^\ast DK_1}(M_B^2,M_B^{'2},Q^2)=-\frac{1}{4\pi^2}\frac{1}{\lambda_{D_{s0}}\lambda_{D}\lambda_{K_1}\mathrm{e}^{-m_{D_{s0}}^2/M_B^2-m_D^2/M_B^{'2}}}(Q^2+m_K^2)\int_{m_c^2}^{s_0}\mathrm{d}s^{'}\int_{m_c^2}^{u_0}\mathrm{d}u^{'}\rho(s^{'},u^{'},t)\;\mathrm{e}^{-s^{'}/M_B^2-u^{'}/M_B^{'2}},
\end{equation}
where
\begin{equation}
\rho(s,u,t)=-\frac{96\pi^4t(m_c^4-m_c^2(s-t+u)+su)}{\lambda^{3/2}+(s,u,t)}+\frac{4\pi^4\GG t}{3\lambda^{3/2}(s,u,t)}.
\end{equation}
\par For $D_{s} DK_0^\ast$ process, the sum rule can be obtained as
\begin{equation}\label{Eq:gDsDK0}
g_{D_{s} DK_0^\ast}(M_B^2,M_B^{'2},Q^2)=-\frac{1}{4\pi^2}\frac{2}{\lambda_{D_{s}}\lambda_{D}\lambda_{K^\ast_0}\mathrm{e}^{-m_{D_{s}}^2/M_B^2-m_D^2/M_B^{'2}}}\frac{Q^2+m_{K_0^\ast}^2}{m_{D_{s}}^2+m_{D}^2+Q^2}\int_{m_c^2}^{s_0}\mathrm{d}s^{'}\int_{m_c^2}^{u_0}\mathrm{d}u^{'}\rho(s^{'},u^{'},t)\;\mathrm{e}^{-s^{'}/M_B^2-u^{'}/M_B^{'2}},
\end{equation}
where
\begin{equation}
\begin{split}
\rho(s,u,t)=&\frac{48\pi^4}{\lambda^{5/2}(s,u,t)}\Bigg(2m_c^3(\lambda(s,u,t)+3t(s-t+u))(m_c^2-m_cm_s-s+t-u)-m_su(s-t+u)(\lambda(s,u,t)+6st)\\
&+m_c\Big(m_cm_s(s^3+s^2(t+u)-5s(t-u)^2+3(t-u)^2(t+u))-u(s(s-2t)(2s-t)+t^3)-u^2(2s+3t)(s+t-u)\\
&+s(s-t)^3+u^4\Big)\Bigg)+\frac{24\pi^4\GG m_c(s(t-2u)+t(u-t))}{\lambda^{5/2}(s,u,t)}.
\end{split}
\end{equation}
\par For $D_{s1} DK_0^\ast$ process, the sum rule can be obtained as
\begin{equation}\label{Eq:gDs1DK0}
g_{D_{s1} DK_0^\ast}(M_B^2,M_B^{'2},Q^2)=-\frac{1}{4\pi^2}\frac{2m_{D_{s1}}^2}{\lambda_{D_{s1}}\lambda_{D}\lambda_{K^\ast_0}\mathrm{e}^{-m_{D_{s1}}^2/M_B^2-m_D^2/M_B^{'2}}}\frac{Q^2+m_{K_0^\ast}^2}{m_{D_{s1}}^2+m_D^2+Q^2}\int_{m_c^2}^{s_0}\mathrm{d}s^{'}\int_{m_c^2}^{u_0}\mathrm{d}u^{'}\rho(s^{'},u^{'},t)\;\mathrm{e}^{-s^{'}/M_B^2-u^{'}/M_B^{'2}},
\end{equation}
where
\begin{equation}
\rho(s,u,t)=-\frac{48\pi^4tu(2m_c^2-s+t-u)}{\lambda^{3/2}(s,u,t)}+\frac{2\pi^4\GG(s-t-3u)}{3\lambda^{3/2}(s,u,t)}.
\end{equation}


\begin{thebibliography}{100}
\bibitem{Gell-Mann:1964ewy}
M. ~Gell-Mann, Phys. Lett. \textbf{8}, 214 (1964)
\bibitem{1964-Zweig-p-}
G.~Zweig, 
\newblock { in: D.Lichtenberg, S.P.Rosen(Eds.), Developments in the Quark
  Theory of Hadrons}, VOL. 1. 1964 - 1978:pp. 22--101, 1964.

\bibitem{ParticleDataGroup:2022pth}
R. L.~Workman \textit{et al.} [Particle Data Group],
PTEP \textbf{2022}, 083C01 (2022)

\bibitem{Nielsen:2009uh}
M.~Nielsen,F.S.~Navarra,S.H.~and Lee,
Phys. Rept. \textbf{497}, 41 (2010)
\bibitem{Chen:2016qju}
H.~X.~Chen, W.~Chen, X.~Liu and S.~L.~Zhu,
Phys. Rept. \textbf{639}, 1-121 (2016)
\bibitem{Richard:2016eis}
J.-M.~Richard, Few Body Syst. \textbf{57}, 1185 (2016)
\bibitem{Esposito:2016noz}
A.~Esposito, A.~Pilloni and A.~D.~Polosa,
Phys. Rept. \textbf{668}, 1-97 (2017)
\bibitem{Ali:2017jda}
A.~Ali, J.S.~ Lange, and S.~Stone, 
Prog. Part. Nucl. Phys. \textbf{97}, 123 (2017)
\bibitem{Guo:2017jvc}
F.~K.~Guo, C.~Hanhart, U.~G.~Mei\ss{}ner, Q.~Wang, Q.~Zhao and B.~S.~Zou,
Rev. Mod. Phys. \textbf{90}, no.1, 015004 (2018)
\bibitem{Albuquerque:2018jkn}
R.M.~Albuquerque, J.M.~ Dias, K.P.~ Khemchandani, A.M.~ Torres, F.S.~ Navarra, M.~ Nielsen, and C.M.~ Zanetti, 
J. Phys. G \textbf{46}, 093002 (2019)

\bibitem{Liu:2019zoy}
Y.~R.~Liu, H.~X.~Chen, W.~Chen, X.~Liu and S.~L.~Zhu,
Prog. Part. Nucl. Phys. \textbf{107}, 237-320 (2019)

\bibitem{Brambilla:2019esw}
N.~Brambilla, S.~Eidelman, C.~Hanhart, A.~Nefediev, C.~P.~Shen, C.~E.~Thomas, A.~Vairo and C.~Z.~Yuan,
Phys. Rept. \textbf{873}, 1-154 (2020)
\bibitem{Richard:2019cmi}
J.-M.~Richard, A.~ Valcarce, and J.~Vijande, 
Annals Phys. \textbf{412}, 168009 (2020)
\bibitem{Faustov:2021hjs}
R.N.~Faustov, V.O.~ Galkin, and E.M.~Savchenko, 
Universe \textbf{7}, 94 (2021)
\bibitem{Chen:2022asf}
H.~X.~Chen, W.~Chen, X.~Liu, Y.~R.~Liu and S.~L.~Zhu,
Rept. Prog. Phys. \textbf{86}, 026201 (2023)

\bibitem{Meng:2022ozq}
L.~Meng, B.~Wang, G.~J.~Wang and S.~L.~Zhu,
Phys.Rept. \textbf{1019} ,1-149 (2023)
\bibitem{D0:2016mwd}
V. M.~Abazov \textit{et al.}[D0],
Phys. Rev. Lett. \textbf{117}, no.2, 022003 (2016)
\bibitem{LHCb:2016dxl}
R.~Aaij \textit{et al.}[LHCb],
Phys. Rev. Lett. \textbf{117} no.15, 152003 (2016)[Addendum: Phys. Rev. Lett. \textbf{118}, no.10, 109904 (2017)]
\bibitem{CMS:2017hfy}
A. M.~Sirunyan \textit{et al.}[CMS],
Phys. Rev. Lett. \textbf{120}, no.20, 202005 (2018)
\bibitem{CDF:2017dwr}
T.~Aaltonen \textit{et al.}[CDF],
Phys. Rev. Lett. \textbf{120}, no.20, 202006 (2018)
\bibitem{ATLAS:2018udc}
M.Aaboud \textit{et al.}[ATLAS],
Phys. Rev. Lett. \textbf{120}, no.20, 202007 (2018) 
\bibitem{LHCb:2020bls}
R.~Aaij \textit{et al.}[LHCb],
Phys. Rev. Lett. \textbf{125}, 242001 (2020)
\bibitem{LHCb:2020pxc}
R.~Aaij \textit{et al.}[LHCb],
Phys. Rev. D \textbf{102}, 112003 (2020)
\bibitem{Yu:2023avh}
Z.~Yu, Q.~Wu, and D.-Y.~Chen, 
arXiv: 2310.12398 [hep-ph]
\bibitem{Cheng:2020nho}
J.-B.~Cheng, S.-Y.~Li, Y.-R.~Liu, Y.-N.~ Liu, Z.-G.~Si, and T.~Yao,
Phys. Rev. D \textbf{101}, no.11, 114017 (2020) 
\bibitem{Guo:2021mja}
T.~Guo, J.~Li, J.~Zhao, and  L.~He,
Phys. Rev. D \textbf{105}, 054018 (2022)
\bibitem{He:2020jna}
X.-G.~He, W.~Wang, and R.~Zhu, 
Eur. Phys. J. C \textbf{80}, no.11, 1026 (2020)
\bibitem{Wang:2020prk}
G.-J.~Wang, L.~Meng, L.-Y.~Xiao, M.~Oka, and S.-L.~Zhu,
Eur. Phys. J. C \textbf{81}, no.2, 188 (2021)
\bibitem{Zhang:2020oze}
J.-R.~Zhang,
Phys. Rev. D \textbf{103}, no.5, 054019 (2021)
\bibitem{Wang:2020xyc}
Z.-G.~Wang,
Int. J. Mod. Phys. A \textbf{35}, no.30, 2050187 (2020)
\bibitem{Lu:2020qmp}
Q.-F.~L\"u, D.-Y.~Chen, and Y.-B.~Dong, 
Phys. Rev. D \textbf{102}, no.7, 074021 (2020)
\bibitem{Agaev:2022eeh}
S.S.~Agaev, K.~Azizi, and H.~Sundu,
Phys. Rev. D \textbf{106}, no.1, 014019 (2022)

\bibitem{Hu:2020mxp}
M.-W.~Hu, X.-Y.~Lao, P.~Ling, and Q.~Wang,
Chin. Phys. C \textbf{45}, no.2, 021003 (2021)
\bibitem{Kong:2021ohg}
S.-Y.~Kong, J.-T.~Zhu, D.~Song, and J.~He,
Phys. Rev. D \textbf{104} no.9, 094012 (2021)
\bibitem{Ke:2022ocs}
H.-W.~Ke, Y.-F.~Shi, X.-H.~Liu, and X.-Q.~Li,
Phys. Rev. D \textbf{106}, no.11, 114032 (2022)
\bibitem{Liu:2020nil}
M.-Z.~Liu, J.-J.~Xie, L.-S.~Geng,
Phys. Rev. D \textbf{102}, no.9, 091502 (2020)
\bibitem{Wang:2021lwy}
B.~Wang, and S.-L.~Zhu,
Eur. Phys. J. C \textbf{82}, no.5, 419 (2022)
\bibitem{Chen:2020aos}
H.-X.~Chen, W.~Chen, R.-R.~Dong, and N.~Su,
Chin. Phys. Lett. \textbf{37}, no.10, 101201 (2020)
\bibitem{Agaev:2020nrc}
S.S.~Agaev, K.~Azizi, and H.~Sundu, 
J. Phys. G \textbf{48}, no.8, 085012 (2021)
\bibitem{Chen:2021erj}
H.-X.~Chen,
Phys. Rev. D \textbf{105}, no.9, 094003 (2022)
\bibitem{Mutuk:2020igv}
H.~Mutuk,
J. Phys. G \textbf{48}, no.5, 055007 (2021)
\bibitem{Molina:2010tx}
R.~Molina, T.~Branz, and E.~Oset,
Phys. Rev. D \textbf{82}, 014010 (2010)
\bibitem{Molina:2020hde}
R.~Molina and E.~Oset,
Phys. Lett. B \textbf{811}, 135870 (2020) [Erratum: Phys. Lett. B \textbf{837}, 137645 (2023)]

\bibitem{Huang:2020ptc}
Y.~Huang, J.-X.~Lu, J.-J.~Xie, and L.-S.~Geng,
Eur. Phys. J. C \textbf{80}, no.10, 973 (2020)
\bibitem{Xiao:2020ltm}
C.-J.~Xiao, D.-Y.~Chen, Y.-B.~Dong, and G.-W.~Meng,
Phys. Rev. D \textbf{103}, no.3, 034004 (2021)
\bibitem{Burns:2020xne}
T. J.~Burns, and E. S.~Swanson, 
Phys. Rev. D \textbf{103}, no.1, 014004 (2021)
\bibitem{Burns:2020epm}
T. J.~Burns, and E. S.~Swanson, 
Phys. Lett. B \textbf{813}, 136057 (2021)
\bibitem{Chen:2020eyu}
Y.-K.~Chen, J.-J.~Han, Q.-F.~L\"u, J.-P.~Wang, and F.-S.~Yu,
Eur. Phys. J. C \textbf{81}, no.1, 71 (2021)

\bibitem{Hsiao:2021tyq}
Y.-K.~Hsiao, and Y.~Yu,
Phys. Rev. D \textbf{104}, no.3, 034008 (2021)
\bibitem{Li:2002pj}
J.-W.~Li, M.-Z.~Yang, and D.-S.~Du, 
HEPNP \textbf{27}, 665-672 (2003)
\bibitem{Cheng:2004ru}
H.-Y.~Cheng, C.-K.~Chua, and A.~Soni, 
Phys. Rev. D \textbf{71}, 014030 (2005)
\bibitem{Lu:2005mx}
C.-D.~L\"u, Y.-L.~Shen, and W.~Wang,
Phys. Rev. D \textbf{73}, 034005 (2006)
\bibitem{Wirbel:1985ji}
M.~Wirbel, B.~Stech, and M.~Bauer,
Z. Phys. C \textbf{29}, 637 (1985)
\bibitem{Bauer:1986bm}
M.~Bauer, B.~Stech, and M.~Wirbel, 
Z. Phys. C \textbf{34}, 103 (1987)
\bibitem{Buchalla:1995vs}
G.~Buchalla, A.J.~Buras, and M.E.~Lautenbacher, 
Rev. Mod. Phys. \textbf{68}, 1125-1144 (1996)
\bibitem{Gortchakov:1995im}
O. Gortchakov, M.P. Locher, V.E. Markushin, and S. von Rotz, 
Z. Phys. A \textbf{353}, 447 (1996)

\bibitem{Cheng:2003sm}
H.-Y.~Cheng, C.-K.~Chua, and C.-W.~Hwang, 
Phys. Rev. D \textbf{69}, 074025 (2004)
\bibitem{Reinders:1984sr}
L.~J. Reinders, H.~Rubinstein, and S.~Yazaki, Phys. Rep. \textbf{127}, 1 (1985)
\bibitem{Shifman:1978bx}
M.~A. Shifman, A.~I. Vainshtein, and V.~I. Zakharov, Nucl. Phys. \textbf{B147}, 385 (1979)
\bibitem{Colangelo:2000dp}
P.~Colangelo and A.~Khodjamirian, 
\textit{At the Frontier of Particle Physics}, edited by M. Shifman (World Scientific, Singapore, 2001), Vol. \textbf{3}, pp. 1495–1576
\bibitem{Narison:2002woh}
S.~Narison,
Camb. Monogr. Part. Phys. Nucl. Phys. Cosmol. \textbf{17}, 1-812 (2007) Cambridge University Press, 2022, ISBN 978-1-00-929029-6, 978-1-00-929031-9, 978-1-00-929033-3, 978-0-521-03731-0, 978-0-521-81164-4, 978-0-511-18948-7

\bibitem{Gelhausen:2013wia}
P.~Gelhausen, A.~Khodjamirian, A.A.~Pivovarov, and D.~Rosenthal, 
Phys. Rev. D \textbf{88}, 014015 (2013) [Erratum: Phys. Rev. D \textbf{91}, 099901 (2015)]
\bibitem{Gubernari:2023rfu}
N.~Gubernari, A.~Khodjamirian, R.~Mandal, and T.~Mannel,
JHEP \textbf{12}, 015 (2023)
\bibitem{tHooft:2008rus}
G.~'t Hooft, G.~Isidori, L.~Maiani, A. D.~Polosa, and V.~Riquer,
Phys. Lett. B \textbf{662}, 424 (2008)

\bibitem{Narison:1989aq}
S.~Narison, QCD spectral sum rules, volume \textbf{26} (1989)
\bibitem{Jamin:2001zr}
M.~Jamin, J.~A. Oller, and A.~Pich, Eur. Phys. J. C \textbf{24}, 237 (2002)
\bibitem{Jamin:1998ra}
M.~Jamin and A.~Pich, Nucl. Phys. B Proc. Suppl. \textbf{74}, 300 (1999)
\bibitem{Ioffe:1981kw}
B.~L. Ioffe, Nucl. Phys. \textbf{B188}, 317 (1981), [Erratum: Nucl.Phys.B \textbf{191}, 591–592 (1981)]
\bibitem{Chung:1984gr}
Y.~Chung, H.~G. Dosch, M.~Kremer, and D.~Schall, Z. Phys. C \textbf{25}, 151 (1984)
\bibitem{Dosch:1988vv}
H.~G. Dosch, M.~Jamin, and S.~Narison, Phys. Lett. B \textbf{220}, 251 (1989)
\bibitem{Khodjamirian:2011ub}
A.~Khodjamirian, T.~Mannel, N.~Offen, and Y.~M. Wang, Phys. Rev. D \textbf{83}, 094031 (2011)
\bibitem{Francis:2018jyb}
A.~Francis, R.~J. Hudspith, R.~Lewis, and K.~Maltman, Phys. Rev. D \textbf{99}, 054505 (2019)
\bibitem{Casalbuoni:1996pg}
R.~Casalbuoni, A.~Deandrea, N.~Di Bartolomeo, R.~Gatto,
F.~Feruglio and G.~Nardulli, Phys. Rept. \textbf{281}, 145 (1997)
\bibitem{Casalbuoni:1992gi}
R.~Casalbuoni, A. Deandrea, N. Di Bartolomeo, R. Gatto,
F. Feruglio and G. Nardulli, Phys. Lett. B \textbf{292}, 371 (1992)
\bibitem{Casalbuoni:1992dx}
R. Casalbuoni, A. Deandrea, N. Di Bartolomeo, R. Gatto,
F. Feruglio and G. Nardulli, Phys. Lett. B \textbf{299}, 139 (1993)

\bibitem{Navarra:2000ji}
F.S.~Navarra, M.~Nielsen, M.E.~Bracco, M.~Chiapparini, and C.L.~Schat,
Phys. Lett. B \textbf{489}, 319 (2000)
\bibitem{Bracco:2006xf}
M.E.~Bracco, A.~Cerqueira Jr., M.~Chiapparini, A.~Loz\'ea, and M.~Nielsen,
Phys. Lett. \textbf{B 641}, 286 (2006)
\bibitem{Colangelo:1995ph}
P.~Colangelo, F. De Fazio, G.~Nardulli, N. Di Bartolomeo, and R.~Gatto,
Phys. Rev. D \textbf{52}, 6422 (1995)
\bibitem{Lin:2024hys}
J.-X.~Lin, H.-X.~Chen, W.-H.~Liang, C.-W.~Xiao, and E.~Oset,
Eur. Phys. J. C \textbf{84}, no.4, 439 (2024)
\bibitem{Janbazi:2017mpb}
M.~Janbazi and R.~Khosravi,
Eur. Phys. J. C \textbf{78}, no.7, 606 (2018)
\bibitem{Momeni:2020scc}
S.~Momeni and R.~Khosravi, arXiv: 2003.04165 
\bibitem{Sundu:2011vz}
H.~Sundu, J.Y.~Sungu, S.~Sahin, N.~Yinelek, and K.~Azizi,
Phys. Rev. D \textbf{83}, 114009 (2011)

\end{thebibliography}
\end{document}